\documentclass[aps,pra,twocolumn,superscriptaddress,reprint,nofootinbib]{revtex4-1}

\usepackage{graphicx}
\usepackage[colorlinks=true,citecolor=blue]{hyperref}
\usepackage{units}
\usepackage{amsfonts,amssymb,amsmath,xcolor}

\newcommand{\llangle}{\langle\!\langle}
\newcommand{\rrangle}{\rangle\!\rangle}

\begin{document}
\title{Thermal transistor and thermometer based on Coulomb-coupled conductors}

\author{Jing Yang}
\affiliation{Department of Physics and Astronomy, University of Rochester, Rochester, New York 14627, USA}
\author{Cyril Elouard}
\affiliation{Department of Physics and Astronomy, University of Rochester, Rochester, New York 14627, USA}
\author{Janine Splettstoesser}
\affiliation{Department of Microtechnology and Nanoscience (MC2), Chalmers University of Technology, S-412 96 G\"oteborg, Sweden}
\author{Bj\"orn Sothmann}
\affiliation{Theoretische Physik, Universit\"at Duisburg-Essen and CENIDE, D-47048 Duisburg, Germany}
\author{Rafael S\'anchez}
\affiliation{Departamento de F\'isica Te\'orica de la Materia Condensada and Condensed Matter Physics Center (IFIMAC), Universidad Aut\'onoma de Madrid, 28049 Madrid, Spain}
\author{Andrew N. Jordan}
\affiliation{Department of Physics and Astronomy, University of Rochester, Rochester, New York 14627, USA}
\affiliation{Institute for Quantum Studies, Chapman University, 1 University Drive, Orange, CA 92866, USA}

\date{\today}

\begin{abstract}
We study a three-terminal setup consisting of a single-level quantum dot capacitively coupled to a quantum point contact. 
The point contact connects to a source and drain reservoirs while the quantum dot is coupled to a single base reservoir. This setup has been used to implement a noninvasive, nanoscale thermometer for the bath reservoir by detecting the current in the quantum point contact. Here, we demonstrate that the device can also be operated as a thermal transistor where the average (charge and heat) current through the quantum point contact is controlled via the temperature of the base reservoir. We characterize the performances of this device both as a transistor and a thermometer, and derive the operating condition maximizing their respective sensitivities.
The present analysis is useful for the control of charge and heat flow and high precision thermometry at the nanoscale.
\end{abstract}

\maketitle

\section{Introduction}

Engineering and controlling heat and its coupling with electricity at the nanoscale is one of the big  challenges that the field of nanoelectronics is facing today. The phenomenon of thermoelectric transport at the mesoscopic level has been widely explored in nanostructures to achieve this goal~\cite{sothmann_thermoelectric_2015,benenti_fundamental_2017}. This includes the design of nanoscale heat engines~\cite{staring_coulomb-blockade_1993,dzurak_observation_1993,humphrey_reversible_2002,entin-wohlman_three-terminal_2010,sanchez_optimal_2011,sothmann_rectification_2012,sothmann_magnon-driven_2012,bergenfeldt_hybrid_2014,sanchez_chiral_2015,hofer_quantum_2015,roche_harvesting_2015,hartmann_voltage_2015,thierschmann_three-terminal_2015,Whitney2016Jan,Schulenborg2017Dec,josefsson_quantum-dot_2018,Sanchez2018Nov}, refrigerators~\cite{giazotto_opportunities_2006,pekola_normal-metal-superconductor_2007,edwards_quantum-dot_1993,prance_electronic_2009,zhang_three-terminal_2015,koski_-chip_2015,hofer_autonomous_2016,sanchez_correlation-induced_2017}, thermal rectifiers~\cite{scheibner_quantum_2008,ruokola_single-electron_2011,fornieri_normal_2014,jiang_phonon_2015,sanchez_heat_2015,martinez-perez_rectification_2015}, and thermal transistors~\cite{li_negative_2006,jiang_phonon_2015,joulain_quantum_2016,sanchez_single-electron_2017,sanchez_all-thermal_2017,zhang_coulomb-coupled_2018,Tang2019Feb,guo2018quantum}. Also the detection of heat flows in such devices via nanoscale low-temperature thermometers~\cite{correa_individual_2015,hofer_quantum_2017,mehboudi_thermometry_2018,de_pasquale_quantum_2018} has been addressed. Remarkable progress has been recently achieved with different mesoscopic devices for milikelvin~\cite{spietz_primary_2003,spietz_shot_2006,gasparinetti2011probing,mavalankar_non-invasive_2013,feshchenko_primary_2013,maradan_gaas_2014,feshchenko_tunnel-junction_2015,iftikhar_primary_2016,Ahmed2018Oct,karimi2018noninvasive,halbertal2016nanoscale} or ultrafast \cite{zgirski2018nanosecond,wang2018fast,brange2018nanoscale}  thermometry.

Recent interest has been raised in designing multiterminal devices able to separate charge and heat currents. Proposals include the use of three terminal configurations of capacitively coupled quantum dots where transport through two (source and drain) terminals responds to charge fluctuations in the third one (the base), involving heat but no charge transfer. Setups of this kind allow for the realization of heat engines~\cite{sanchez_optimal_2011,sothmann_rectification_2012,roche_harvesting_2015,hartmann_voltage_2015,thierschmann_three-terminal_2015,Whitney2016Jan,dare_powerful_2017,walldorf_thermoelectrics_2017,strasberg_fermionic_2018}, refrigerators~\cite{zhang_three-terminal_2015,koski_-chip_2015,sanchez_correlation-induced_2017,erdman_absorption_2018}, thermal transistors~\cite{thierschmann_thermal_2015,sanchez_single-electron_2017,sanchez_all-thermal_2017,zhang_coulomb-coupled_2018} and thermometers~\cite{Zhang2018Nov}. Of particular interest is the case of thermal transistors and non-invasive thermometers, where one seeks to maximize the response of the system (a current from source to drain) by minimizing the injection of heat from the base terminal. In the first case, transport in the system is modulated by changes in the base temperature, $\Theta$. Conversely in the second case, the current serves as readout of the temperature of the base. However, the tunneling current through a weakly-coupled quantum dot (based on single-electron transitions) is small.

\begin{figure}[b]
\includegraphics[width=\linewidth]{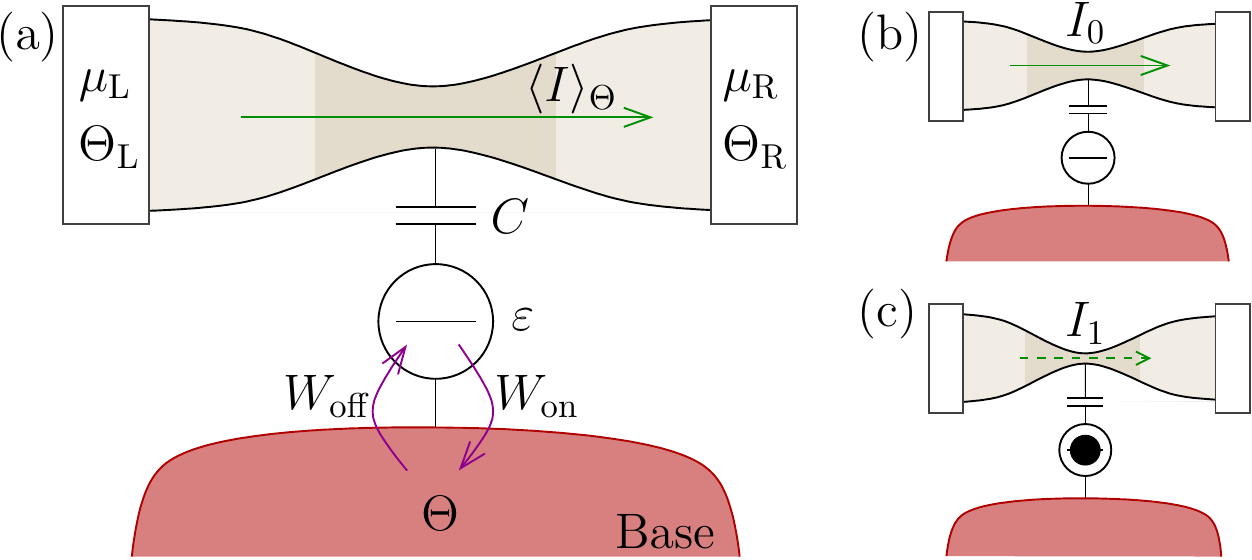}
\caption{\label{fig:three-terminal-device} (a) The QPC (beige region) and a single electron quantum dot connected to three terminals. Source and drain reservoirs have chemical potentials $\mu_\text{L},\mu_\text{R}$ and temperatures $\Theta_\text{L},\Theta_\text{R}$. The base reservoir, with temperature $\Theta$ and electrochemical potential $\mu$, is tunnel-coupled to the quantum dot with tunneling rates $W_{\rm on/off}$. The capacitance $C$ mediates the dependence of the current through the QPC, $I_\alpha$, upon the charge state $\alpha$ of the quantum dot, (b) $\alpha=0$ or (c) $\alpha=1$.}
\end{figure}
To overcome this issue, we consider an alternative structure, shown in Fig.~\ref{fig:three-terminal-device}, consisting of a capacitively coupled quantum dot and a quantum point contact (QPC). The quantum dot is tunnel coupled to the base reservoir, which has a given temperature $\Theta$.
The steady-state population of the quantum dot impacts the average (charge and heat) currents through the QPC, which in turn is connected to source and drain reservoirs.  This structure has been employed previously to study the full counting statistics of single electron transport in quantum dots both theoretically~\cite{korotkov_continuous_1999,korotkov_selective_2001,korotkov_noisy_2002,pilgram_efficiency_2002,jordan_continuous_2005,jordan_quantum_2005,jordan_qubit_2006,jordan_leggett-garg_2006,sukhorukov_conditional_2007,flindt_universal_2009} and experimentally~\cite{gustavsson_counting_2006,fujisawa_bidirectional_2006,ubbelohde_measurement_2012,kung_irreversibility_2012,hofmann_equilibrium_2016,hofmann_measuring_2016,entin-wohlman_heat_2017}, where the QPC acts as a charge detector that monitors the occupation of the dot. This setup was also experimentally demonstrated \cite{gasparinetti2012nongalvanic,torresani_nongalvanic_2013,mavalankar_non-invasive_2013,maradan_gaas_2014} to behave as a thermometer, enabled by the fact that the population of the dot is sensitive to the temperature in the base reservoir.

In this paper, we show that the structure can be used as a thermal transistor as well. We analyze its performance both as a transistor and a thermometer by deriving the corresponding sensitivities and finding the operating conditions optimizing them. When the device is operated as a thermal transistor, the goal is that a small temperature change in the base reservoir triggers a large change of the average charge or heat current in the QPC. We quantify the device performance by the differential sensitivity of the average charge and heat current to the infinitesimal temperature change in the base, as well as by the power gain. 

For the operation of the device as a thermometer, we apply metrological tools to characterize two measurement protocols. The first one involves the sequential coupling to the dot to the probed reservoir and the QPC. This corresponds to the paradigmatic protocol of classical and quantum metrology~\cite{giovannetti_advances_2011}. In the second protocol, both interactions are always turned on. This latter procedure is easier to implement and was actually used in Refs.~\cite{gasparinetti2012nongalvanic,torresani_nongalvanic_2013,mavalankar_non-invasive_2013,maradan_gaas_2014}.
Both protocols allow noninvasive temperature measurements since single electron tunneling only involves a very small amount of energy and charge exchange between the measured reservoir (the base) and part of the thermometer (the quantum dot). 
When optimally operated, we find that the thermometer's sensitivity is for both protocols limited by telegraph noise induced in the QPC by electron tunneling in the dot. 
Interestingly, the optimal sensitivity of the thermometer occurs for the same quantum-dot parameters as the optimal sensitivity of the thermal transistor. In contrast to the thermal transistor, we find that  the sensitivity limits of the thermometer do not depend on the QPC current in the two quantum-dot charge configurations. For the thermal transistor, instead, we find that the power gain is independent of the dot occupation and the base-reservoir temperature.

With respect to the previously studied setup containing two capacitively coupled dots, the device shown in Fig.~\ref{fig:three-terminal-device} has two definite advantages. First, the average currents flowing through the QPC are much larger than in the quantum-dot case where the Coulomb-blockade much reduces the conductance compared to the conductance quantum~\cite{nazarov_quantum_2009}. Furthermore, backaction is suppressed: In a QPC, there is no significant charge-buildup in the vicinity of the saddle point potential, which forms the QPC. As a consequence, the quantum dot state modifies the transmission probability of the QPC without energy exchange between the quantum dot and the QPC. This property allows to obtain high power gain for the transistor and participates to make the thermometer noninvasive as only a small bounded amount of energy flows back and forth between the dot and the base without involving the QPC. This situation is different for transport through a quantum dot connecting source and drain lead. In the latter case, random fluctuations of the electrostatic potential caused by the nearby, capacitively coupled gate dot~\cite{sanchez_all-thermal_2017,sanchez_single-electron_2017} (or simply environmental fluctuations~\cite{sothmann_rectification_2012,ruokola_theory_2012,rossello_dynamical_2017,entin-wohlman_heat_2017}) do induce energy exchange between the source-drain system and the base part in general. Exceptions have been identified involving particular tunneling rate configurations or strongly coupled dots \cite{sanchez_all-thermal_2017,sanchez_single-electron_2017}.

The paper is organized as follows. In Sec.~\ref{sec:setup} we introduce the model of our setup. The operation as a thermal transistor is discussed in Sec.~\ref{sec:thermal-transistor} while the thermometer configuration is analyzed in Sec.~\ref{sec:thermometer}. Our results are summarized and conclusions are drawn in Sec.~\ref{sec:conclusion}.

\section{\label{sec:setup}Setup}

We consider a three-terminal device as shown in Fig.~\ref{fig:three-terminal-device}. It contains a spinless single-level quantum dot, which can be either empty, denoted as $0$, or occupied with a single electron, denoted as $1$ (this corresponds to a quantum dot in the spin-split Coulomb-blockaded regime).  The addition energy $\varepsilon$ is defined as the energy difference of these two states. The dot is weakly tunnel coupled to the base reservoir with chemical potential $\mu$ and temperature $\Theta$. In what follows we choose the electrochemical potential of the base reservoir as a reference energy, $\mu=0$. Throughout this article, we set $k_\text{B} \equiv 1$. The coupling strength between dot and base reservoir is characterized by the rate $\Gamma$. 
The population dynamics of the weakly coupled dot, with $\hbar \Gamma/\Theta\ll 1$ is described
by the rate equations
\begin{eqnarray}
\dot{P}_{0}(t)  = -W_{\text{off}}P_{0}(t)+W_{\text{on}}P_{1}(t)
\label{eq:diff-dot}
\end{eqnarray}
and $\dot{P}_{1}(t)=-\dot{P}_{0}(t)$, where $P_{0}(t)$ and $P_{1}(t)$ denote the probability to find the dot empty or singly occupied at time $t$, respectively. According to Fermi's golden rule, the transition rate from dot state $0$ to $1$ is $W_{\text{off}}=\Gamma f_{\Theta}(\varepsilon)$ and the transition rate from dot state $1$ to $0$ is $W_{\text{on}}=\Gamma[1-f_{\Theta}(\varepsilon)],$ where $f_{\Theta}(\varepsilon)=1/(e^{\varepsilon/\Theta}+1)$ is the Fermi function.
The dot occupations relax to the steady state on a time scale $\Gamma^{-1}$. The steady state populations are
\begin{eqnarray}
P_{0} & = & W_{\text{on}}/\Gamma=1-f_{\Theta}(\varepsilon),\label{eq:P0}\\
P_{1} & = & W_{\text{off}}/\Gamma=f_{\Theta}(\varepsilon).\label{eq:P1}
\end{eqnarray}

The mesoscopic QPC is connected to two macroscopic source and drain reservoirs with electrochemical potentials and temperatures $\mu_\text{L}$, $\Theta_\text{L}$ and $\mu_\text{R}$, $\Theta_\text{R}$ respectively, as shown in Fig. \ref{fig:three-terminal-device}. The average current in the QPC sensitively depends on the state of the nearby quantum dot due to Coulomb interactions. For an ideal, saddle shaped potential landscape where transport occurs only via the lowest transverse mode, the QPC transmission takes the form~\cite{buttiker_quantized_1990}
\begin{equation}
\mathcal{T}_{\alpha}(E)=\frac{1}{1+\exp\left(-2\pi\frac{E-U_{\alpha}}{\hbar\omega}\right) }.
\end{equation}
Here $\alpha=0,\,1$ represents the empty or occupied state of the dot, $\omega$ characterizes the curvature of the potential, and $U_{\alpha}$ is the electrostatic potential at the bottom of the saddle point when the dot is empty or occupied. In the nonlinear transport regime, it is important to account for the dependence of $U_{\alpha}$ on the electrochemical potentials of the leads, $U_{\alpha}=U_{\alpha,0}+\lambda \mu_\text{L}+(1-\lambda)\mu_\text{R}$ with $0\leq\lambda \leq1$, to ensure gauge invariance~\cite{christen_gauge-invariant_1996}.
Throughout this paper, we assume that $\Theta_{i}$ $(i=\text{L,R})$ is much smaller than all relevant energy scales in the QPC subsystem, in particular, $|U_\alpha-\mu_{i}|$  and $eV=\mu_\text{L}-\mu_\text{R}$, so that $f_{\Theta_{i}}(E-\mu_{i})$ can be approximated by the Heaviside function. Then the energy relevant to electronic and heat transport falls in the range $[\mu_\text{R},\,\mu_\text{L}]$. The average charge and heat currents flowing \textit{out of} a reservoir, for a given dot state, can be calculated from the Landauer-B\"uttiker formula~\cite{datta_electronic_1997,blanter_shot_2000}. Focusing on currents flowing out of the left reservoir, we find
\begin{equation}
\langle I_{\alpha}\rangle=\frac{2e}{h}\int_{\mu_\text{R}}^{\mu_\text{L}}dE\mathcal{T}_{\alpha}(E)=\frac{e\omega}{2\pi^{2}}\ln\left[\frac{1+e^{2\pi(\mu_\text{L}-U_{\alpha})/(\hbar\omega)}}{1+e^{2\pi(\mu_\text{R}-U_{\alpha})/(\hbar\omega)}}\right],\label{eq:Ialph-ave}
\end{equation}
\begin{equation}
\langle I_{\alpha}^{Q}\rangle=\frac{2}{h}\int_{\mu_\text{R}}^{\mu_\text{L}}dE\mathcal{T}_{\alpha}(E)(E-\mu_\text{L}).\label{eq:IalphQ-ave}
\end{equation}
where for simplicity, we have chosen not to indicate the reservoir L, where we always assume currents to be detected. The subscript $\alpha$ stands for the dot state.
Given the dot stays in state $\alpha$, the current noise in the QPC is fully characterized by shot noise. Since we assume the low-temperature limit, thermal noise can be neglected. The zero frequency shot noise power spectral density of the charge current---considering auto-correlations in reservoir L---is~\cite{blanter_shot_2000}
\begin{eqnarray}
S_{\alpha} &=& \frac{e^{2}}{\pi\hbar}\int_{\mu_\text{R}}^{\mu_\text{L}}dE\ \mathcal{T}_{\alpha}(E)[1-\mathcal{T}_{\alpha}(E)]\nonumber\\
&=&\frac{e^{2}\omega}{2\pi}\frac{\sinh\left[\frac{\pi(\mu_\text{L}-\mu_\text{R})}{\hbar\omega}\right]}{\cosh\left[\frac{\pi(\mu_\text{L}-U_{\alpha})}{\hbar\omega}\right]\cosh\left[\frac{\pi(\mu_\text{R}-U_{\alpha})}{\hbar\omega}\right]},\label{eq:Salpha}
\end{eqnarray}
where, again, the subscript $\alpha$ indicates the dot state.

\section{\label{sec:thermal-transistor}Thermal transistor}

\subsection{Sensitivity}

\begin{figure}
\begin{centering}
\includegraphics[scale=0.52]{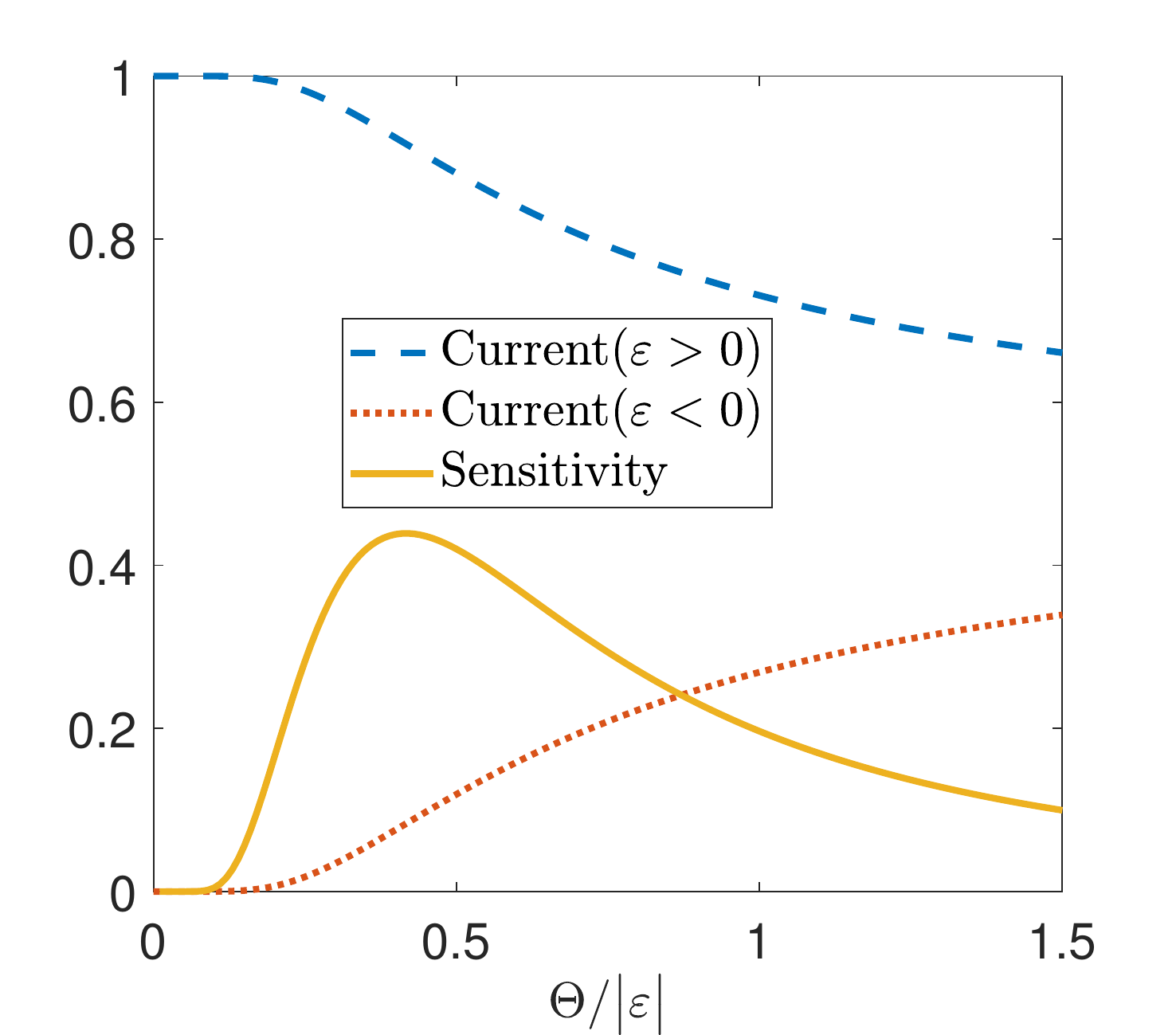}
\par\end{centering}
\caption{\label{fig:current-sens}The normalized current $(\langle I\rangle_{\Theta}-\langle I_{1}\rangle)/\Delta I$ and the normalized differential sensitivity $(\big|\varepsilon\big|/\Delta I)\big|d\langle I\rangle_{\Theta}/d\Theta\big|=\xi(\varepsilon/\Theta)$ versus the normalized temperature $\Theta/\big|\varepsilon\big|$. The blue and the red lines are for the cases $\varepsilon>0$ and $\varepsilon<0$ respectively. }
\end{figure}

When the device under study acts as a thermal transistor in the steady state, analogous to the electric transistor, we would like to see a large change of average charge or heat current in the QPC due to a small change in the temperature of the base reservoir.  Control of charge and heat currents via temperature gradients can be useful if, e.g., a certain device operation should only be performed as long as a certain temperature is not exceeded.

When the dot reaches its stationary state at a given base-reservoir temperature $\Theta$, the average charge current flowing in the QPC is
\begin{equation}
\langle I\rangle_{\Theta}=P_{0}\langle I_{0}\rangle+P_{1}\langle I_{1}\rangle=\langle I_{1}\rangle+[1-f_{\Theta}(\varepsilon)]\Delta I,\label{eq:I-ave}
\end{equation}
where  
\begin{equation}
\Delta I\equiv\langle I_{0}\rangle-\langle I_{1}\rangle.\label{eq:DeltaI}
\end{equation}
For the average heat current one just needs to replace $\langle I\rangle_{\Theta}$, $\langle I_{i}\rangle$ and $\Delta I$ with $\langle I^{Q}\rangle_{\Theta}$, $\langle I_{\alpha}^{Q}\rangle$ and $\Delta I^{Q}\equiv\big|\langle I_{0}^{Q}\rangle-\langle I_{1}^{Q}\rangle\big|$ respectively. We see from Eq.~(\ref{eq:I-ave}) that both the steady-state average charge and heat current depend on the temperature of the base reservoir through the steady-state population of the dot. The differential sensitivity of the average charge current in the QPC to the temperature of the base reservoir is\footnote{Note that we exclude the case $\varepsilon=0$, where the device is fully insensitive to temperature changes.}
\begin{equation}
\Bigg|\frac{d\langle I\rangle_{\Theta}}{d\Theta}\Bigg|=\Bigg|\frac{df_{\Theta}(\varepsilon)}{d\Theta}\Bigg|\Delta I=\xi\left(\varepsilon/\Theta\right)\frac{\Delta I}{\big|\varepsilon\big|},\label{eq:dIdTheta}
\end{equation}
where we have introduced the function for the normalized differential sensitivity
\begin{equation}
\xi(x)\equiv\frac{x^{2}}{2[1+\cosh(x)]}.
\end{equation}
We observe that the asymmetric function $\xi(x)$ reaches its maximum at $\big|x\big|=2.4$ and decreases to half of its value at $\big|x\big|=1$ and $\big|x\big|=4.5$. Thus for fixed $\varepsilon$, the prefactor on the right hand side of Eq.~(\ref{eq:dIdTheta}) reaches its maximum $0.44$ at $\Theta\approx0.4\big|\varepsilon\big|$, with left width $0.2\big|\varepsilon\big|$ and right width $0.6\big|\varepsilon\big|$,  as shown in Fig. \ref{fig:current-sens}. From this, we derive the maximum differential sensitivity as  
\begin{equation}
\max_{\Theta}\Bigg|\frac{d\langle I\rangle_{\Theta}}{d\Theta}\Bigg|=\frac{0.44\Delta I}{\big|\varepsilon\big|}.\label{eq:trans-max-sens}
\end{equation}
For the differential sensitivity of the average heat current, one just needs to replace $\Delta I$ in Eqs. (\ref{eq:dIdTheta}, \ref{eq:trans-max-sens}) with $\Delta I^{Q}$.

\subsection{Power gain}

\begin{figure}[t]
\begin{centering}
\includegraphics[scale=0.5]{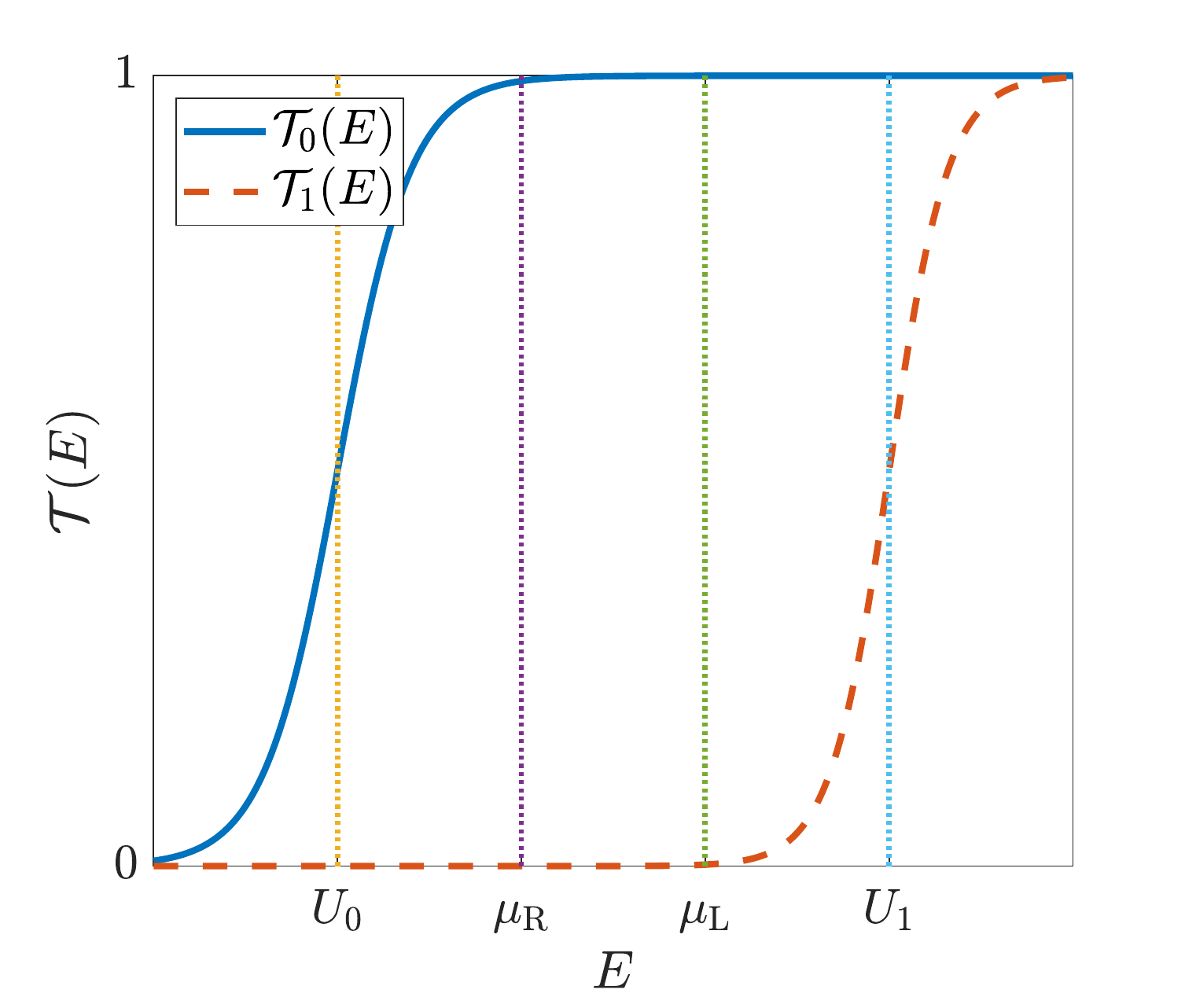}
\par\end{centering}
\caption{\label{fig:MaxDI}QPC transmission in the regime where the different dot occupations correspond to fully closed and open channels. The requirement to reach this regime is to have $(\mu_\text{R}-U_{0})\gg \hbar\omega$ and $(U_{1}-\mu_\text{L})\gg \hbar\omega$.}
\end{figure}

In general, the power gain of a transistor is defined as the ratio between output and input power $\mathcal{G}_{P}\equiv P_{\text{out}}/P_{\text{in}}$. Here, the output power is given by the power difference as the temperature is changed from $\Theta_{1}$ to $\Theta_{2}$ at a given applied voltage $V$, i.e.,
\begin{equation}
P_{\text{out}}=(\langle I\rangle_{\Theta_{2}}-\langle I\rangle_{\Theta_{1}})V=\Delta I\big|f_{\Theta_{2}}(\varepsilon)-f_{\Theta_{1}}(\varepsilon)\big|V.
\end{equation}
The definition of the input power is less straightforward, since within the considered model no energy flows from the quantum dot into the QPC circuit. Instead, the relevant energy flow is here the energy transferred from the base reservoir into the quantum dot within a time duration $\Gamma^{-1}$ set by the characteristic time scale of the quantum-dot tunneling dynamics, when the reservoir temperature is changed from an initial value $\Theta_{1}$ to a final value $\Theta_{2}$. Therefore, we find 
\begin{equation}
P_{\text{in}}=\Gamma \varepsilon\big|f_{\Theta_{2}}(\varepsilon)-f_{\Theta_{1}}(\varepsilon)\big|.
\end{equation}
 With this the result for the power gain is found to be
\begin{equation}
\mathcal{G}_{P}=\frac{V\Delta I}{\Gamma \varepsilon}.\label{eq:Gp}
\end{equation}
 Interestingly, the information about the occupation of the dot and about the temperature of the base reservoir drops out. 

For fixed applied bias $V$, one can further maximize the sensitivity, Eq.~(\ref{eq:trans-max-sens}), and the power gain, Eq.~(\ref{eq:Gp}), over $\Delta I$ or $\Delta I^{Q}$. It is readily found from Eqs.~(\ref{eq:Ialph-ave}) and (\ref{eq:IalphQ-ave}) that the maxima of both quantities are reached when the different dot occupations (empty or occupied) result into a completely closed or open QPC channel, corresponding to $\mathcal{T}_{0}(E)\approx1$ and $\mathcal{T}_{1}(E)\approx0$ respectively. This regime can be reached by tuning the parameters $U_{\alpha}$ and $\omega$ such that one has $(\mu_\text{R}-U_{0})/\hbar\omega\gg1$ and $(U_{1}-\mu_\text{L})/\hbar\omega\gg1$ as shown in Fig.~\ref{fig:MaxDI}. In this regime, we have $\langle I_{1}\rangle=\langle I_{1}^{Q}\rangle=0$, but $\langle I_{0}\rangle=2e(\mu_\text{L}-\mu_\text{R})/h$, and $\langle I_{0}^{Q}\rangle=-2(\mu_\text{L}-\mu_\text{R})^{2}/h$.

\section{\label{sec:thermometer}Thermometer}

\begin{figure}
\begin{centering}
\includegraphics[width=\linewidth]{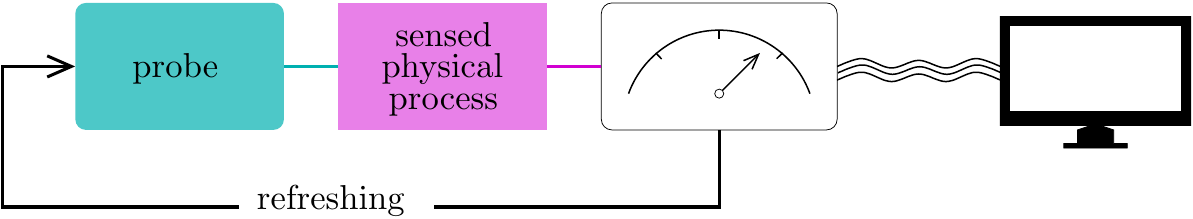}
\par\end{centering}
\caption{\label{fig:The-general-protocol} A typical cycle of a metrological process consists of the following steps: (i) The probe (blue rectangle) is initially prepared in some known state. (ii) The probe interacts with some physical process (magenta rectangle) that depends on the true value of the estimation parameter of interest. The interaction is turned off before moving to the next step. (iii) The interaction between the probe and the measuring apparatus is turned on to perform a noiseless projective measurement. When the measurement is done, the interaction is turned off and the measurement outcome is sent to the computer for later processing. 
}
\end{figure}

We now turn to the operation of the setup as a thermometer. We analyze two different types of protocols with the aim to sense the temperature $\Theta$ of the base reservoir. The standard protocol of metrology, shown in Fig. \ref{fig:The-general-protocol} and described in detail in the figure caption, corresponds to a sequence of discrete measurements: a physical system (the quantum dot, in our case) is used as a probe which is prepared in some known initial state and then undergoes some physical process (charging/uncharging) which depends on the true value of the estimation parameter of interest (the temperature of the base reservoir, $\Theta$). In this way, the information about the estimation parameter is encoded in the state of the probe. Finally, one uses some measuring apparatus (the QPC) to measure the state of the probe. When repeating the above procedure $N$ times, the standard precision of the measurement scales as $1/\sqrt{N}$. An alternative but more practical protocol (this is the one actually used in experiments Refs.~\cite{mavalankar_non-invasive_2013,maradan_gaas_2014}) is to keep both interactions always on, avoiding the sequential coupling and decoupling procedures. 

The QPC as a measurement apparatus or detector of the state of the dot has been widely discussed in the context of mesoscopic measurement processes~\cite{korotkov_continuous_1999,korotkov_selective_2001,korotkov_noisy_2002,pilgram_efficiency_2002,jordan_continuous_2005,jordan_quantum_2005,jordan_qubit_2006,jordan_leggett-garg_2006,sukhorukov_conditional_2007,flindt_universal_2009,gustavsson_counting_2006,fujisawa_bidirectional_2006,ubbelohde_measurement_2012,kung_irreversibility_2012,hofmann_equilibrium_2016,hofmann_measuring_2016,entin-wohlman_heat_2017}. Let us first for simplicity assume that the quantum dot is fixed to be either empty or filled, and that we want to determine the state of the dot by looking at the current flowing in the QPC. Although the on and off states of the dot correspond to different average current $\langle I_{1}\rangle$ and $\langle I_{0}\rangle$ in the QPC, we can not resolve  these two current levels instantaneously, due to the shot noise. Therefore, to measure the state of the dot, one has to switch on the QPC circuit for some time duration. Suppose we measure for this time duration $\tau_\alpha$, which is much longer than the correlation time of the shot noise in the QPC and in principle depends on the dot state $\alpha$. Due to the central limit theorem, the distribution of the time average current $(1/\tau_\alpha)\int I_\alpha(t')dt'$ conditioned on the dot state $\alpha$ is a Gaussian with mean $\langle I_{\alpha}\rangle$ and variance $S_{\alpha}/\tau$, where $S_{\alpha}$ is the zero frequency conditioned shot noise spectral density defined in Eq.~(\ref{eq:Salpha}). To reach a signal to noise ratio of at least unity for distinguishing the two Gaussians, the measurement must be turned on for at least a time duration of $\max\{\tau_0,\tau_1\}$, where $\Delta I$ is defined in Eq.~(\ref{eq:DeltaI}) and
\begin{equation}
\tau_{\alpha}\equiv\frac{S_{\alpha}}{(\Delta I)^{2}}\label{eq:tau-alph}
\end{equation}
is the measurement time. If the QPC circuit is switched on for a time duration much longer than $\max\{\tau_0,\tau_1\}$, it effectively performs an ideal measurement of the quantum-dot occupation.

\subsection{\label{subsec:Standard-protocol}Standard (discrete) protocol}

We now discuss the three steps of the protocol of Fig.~\ref{fig:The-general-protocol}.  Therefore, (i) we initially prepare the quantum dot---the probe of our setup---in state $0$, meaning that no excess electrons are in the dot. Next, (ii) we turn on the interaction with the base reservoir for a time duration $\Delta t^{(ii)}=c^{(ii)}\Gamma^{-1}$, where the constant $c^{(ii)}\gg1$ is sufficiently large to guarantee that the dot reaches the stationary state. We then (iii) turn off the interaction with the base reservoir and at the same time turn on the measurement by the QPC to determine whether there is an electron on the dot or not. We measure for time duration  $\Delta t^{(iii)}=c^{(iii)}\max\{\tau_0,\tau_1\}$ to perform an effectively noiseless ideal occupation measurement, where $c^{(iii)}\gg1$ is chosen to obtain a sufficiently large signal-to-noise ratio here. The measurement gives a binary outcome statistically described by the probability mass function $\{P_{\alpha}\}$, Eqs.~(\ref{eq:P0},\ref{eq:P1}), corresponding to the two outcomes $\alpha=0,\,1$. The corresponding sensitivity of the measurement about the temperature $\Theta$ associated with a single cycle is quantified by the Fisher information, which is given by
\begin{equation}
F_{\Theta}=\sum_{\alpha=0,1}\frac{(\partial_{\Theta}P_{\alpha})^{2}}{P_{\alpha}}=\frac{\xi\left(\varepsilon/\Theta\right)}{\Theta^{2}}.\label{eq:FTheta}
\end{equation}
Note that the inverse of the Fisher information sets the lower bound of the variance of any temperature estimator, known as the Cram\'er-Rao bound~\cite{kay_fundamentals_1993}. 

\begin{figure}
\centering{}\includegraphics[scale=0.43]{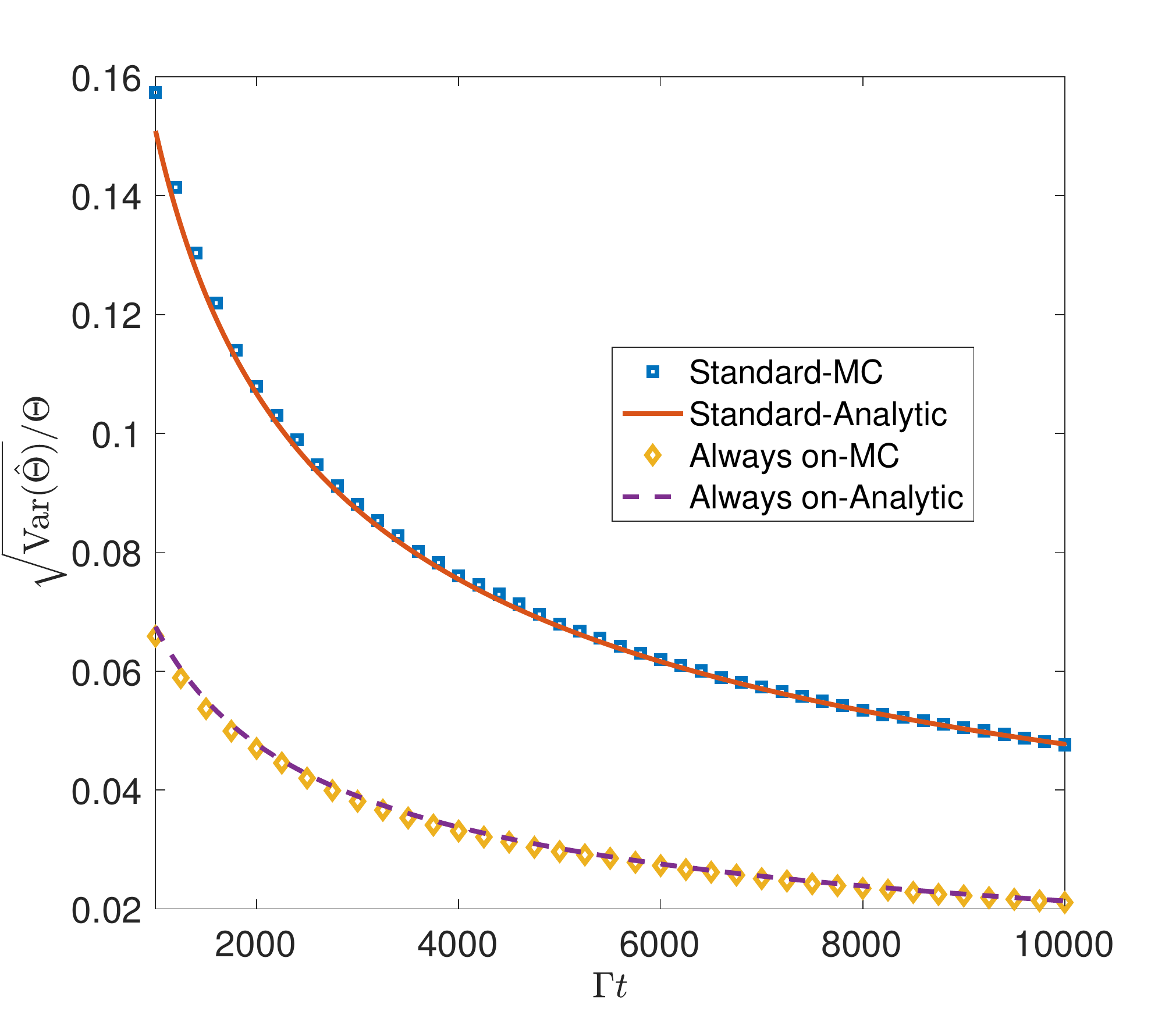}
\caption{\label{fig:MC}Comparison of analytics of the standard, Eq.~(\ref{eq:std-min-var}), and the always-on protocol, Eq.~(\ref{eq:always-min-var}), with Monte Carlo (MC) simulations. We ignore the shot noise in the QPC for both protocols. We take $\varepsilon/\Theta$ to be the optimal value $2.4$ for both protocols. In general, if our prior knowledge about $\Theta$ is very loose and $\varepsilon$ can be far detuned from its optimal value, then the optimal sensitivity shown here can be achieved by adaptive measurements.}
\end{figure}

In a given time $t$, one can repeat the cycle $N=t/(\Delta t^{(ii)}+\Delta t^{(iii)})$ times. We denote the measurement outcome for the $n$-th cycle as $x_{n}$, where $x_{n}=0$ if the dot is empty and $x_{n}=1$ if the dot is occupied. Based on a series of measurement outcomes $\{x_{n}\}$, we propose to apply the asymptotically unbiased maximum likelihood estimator $\hat{\Theta}_{\text{MLE}}(\{x_{n}\})=\varepsilon/\ln[1/\hat{f}(\{x_{n}\})-1]$ to estimate the temperature, where $\hat{f}(\{x_{n}\})=\sum_{n=1}^{N}x_{n}/N$. When $N$ is sufficiently large, the maximum likelihood estimator can saturate the Cram\'er-Rao bound asymptotically~\cite{kay_fundamentals_1993}, which is given by
\begin{eqnarray}
\text{Var}(\hat{\Theta}_{\text{MLE}}) & = & \frac{1}{NF_{\Theta}} \nonumber\\
&=& \frac{c}{t}\frac{\Gamma^{-1}+\max(\tau_{\alpha})}{F_{\Theta}}.\label{eq:VarTemp-noApprox}
\end{eqnarray}
For simplicity, we here assumed $c^{(iii)}=c^{(ii)}\equiv c$. In a typical experiment, one can resolve the state of the dot in a time scale much shorter than the life time of the state limited by electron tunneling. This indicates that the time scale of the tunneling $\Gamma^{-1}$ is much longer than the time scale required to resolve the two current levels, $\max\{\tau_0,\tau_1\}$. With this approximation, Eq.~(\ref{eq:VarTemp-noApprox}) reduces to
\begin{equation}
\text{Var}(\hat{\Theta}_{\text{MLE}})=\frac{c\Theta^{2}}{\Gamma t\xi(\varepsilon/\Theta)},\label{eq:std-var}
\end{equation}
which is independent of the QPC parameters. Now, from the property of the function $\xi(x)$ discussed in Sec. \ref{sec:thermal-transistor},
we see that
\begin{equation}
[\text{Var}(\hat{\Theta}_{\text{MLE}})]_{\min}=\frac{2.3c\Theta^{2}}{\Gamma t},\label{eq:std-min-var}
\end{equation}
and the variance ranges between $[1,\,2]\times$ the minimum variance when $\big|\varepsilon\big|$ is tuned between $[\Theta,\,4.5\Theta]$.
Equation (\ref{eq:std-min-var}) is confirmed by Monte Carlo simulation as shown in Fig.~\ref{fig:MC}, where we generate the noiseless measurement signals (electric current) in the QPC according to the probability mass function $\{P_{\alpha}(t)\}$ described by Eq.~(\ref{eq:diff-dot}). The simulation shows that one needs to take $c\gtrsim10$ in order to obtain a numerical variance (square markers) of the maximum likelihood estimator, which approaches the Cram\'er-Rao bound (\ref{eq:std-min-var}) (red solid line) in the long time limit. 

Since the optimal value of $\big|\varepsilon\big|$ depends on the true value of the estimated temperature, the maximum sensitivity given by Eq.~(\ref{eq:std-min-var}) can only be reached by adaptive measurements consisting of multiple rounds, where one needs to adjust the value of $\big|\varepsilon\big|$ according to the temperature estimate from the previous round. If $\big|\varepsilon\big|$ is kept fixed all the time, as shown in Fig.~\ref{fig:sens-temp}, the normalized standard deviation $\sqrt{\mathrm{Var}(\hat{\Theta})}/\Theta$ of the temperature estimator diverges exponentially at low temperature and linearly at high temperature.

\subsection{Always-on estimation}\label{sec_always_on}

\begin{figure}
\begin{centering}
\includegraphics[scale=0.58]{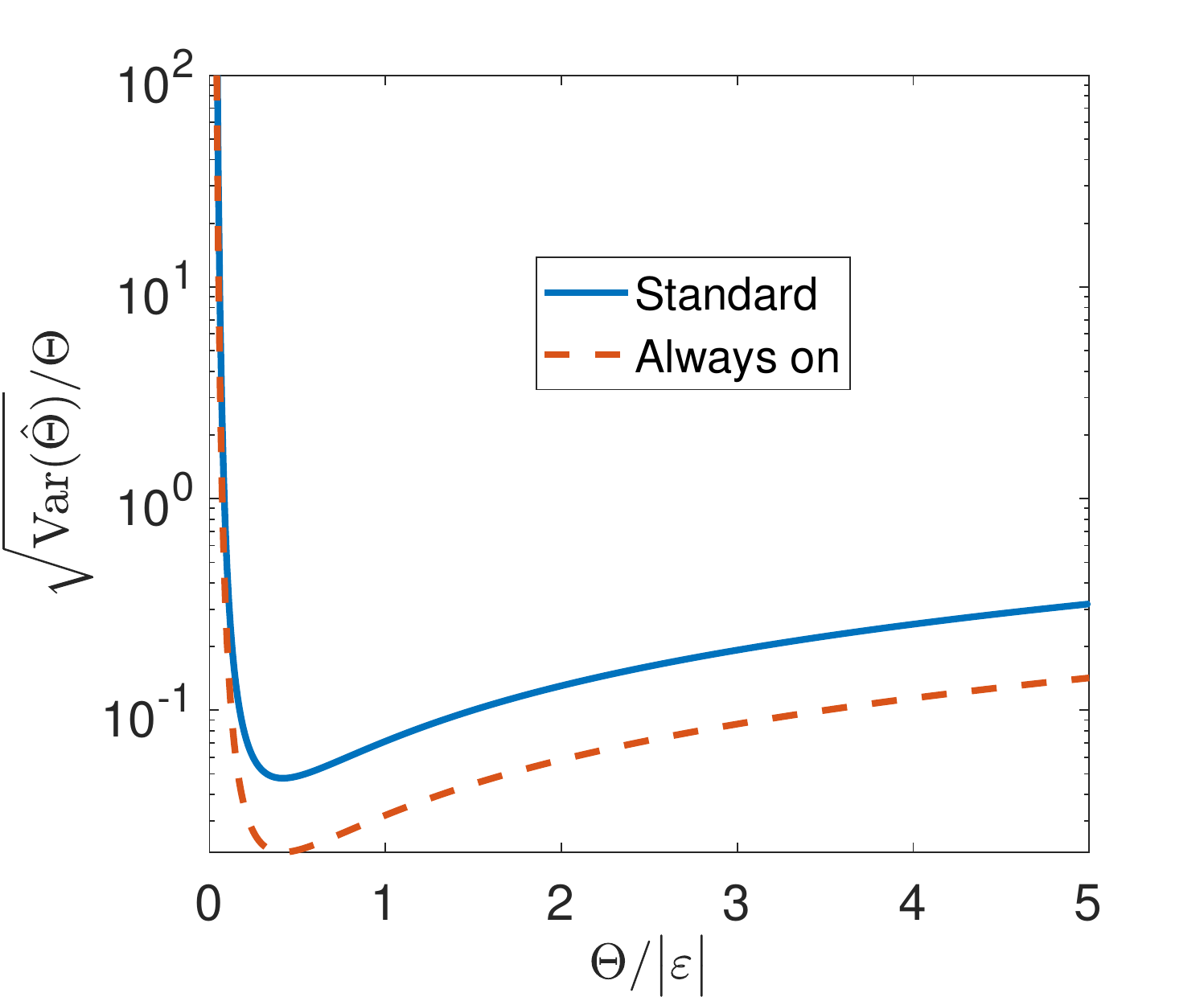}
\par\end{centering}
\caption{\label{fig:sens-temp}The normalized standard deviation of the thermometer against the normalized temperature when  $\big|\varepsilon\big|$ is kept fixed. The blue and red lines are plotted according to Eqs. (\ref{eq:std-var},\ref{eq:always-var}) respectively, where $c=10$ in Eq.(\ref{eq:std-var}). We see that for both protocols, the normalized standard deviation $\sqrt{\mathrm{Var}(\hat{\Theta})}/\Theta$ scales as $\exp(\big|\varepsilon \big|/\Theta)$ at low temperature and $\Theta/\big| \varepsilon \big|$ at high temperature. }
\end{figure}

Let us now consider the alternative protocol, where the measurement by the QPC and the interaction with the electron reservoir are always turned on. We estimate the temperature of the base reservoir using the transferred charge in the QPC. When the dot reaches its steady state, the average current in the QPC is described by Eq.~(\ref{eq:I-ave}), where the impact of the quantum dot enters via its steady-state occupation probabilities. This QPC current has statistical fluctuations, which can be attributed to two different sources of noise. The first is the already mentioned shot noise, described by Eq.~(\ref{eq:Salpha}), due to the partially open channel in the QPC. The other source is telegraph noise, which stems from stochastic switching of the dot states caused by electrons tunneling on and off the quantum dot. We define the transferred charge in the QPC between the beginning of the measurement at $t'=0$ up to time $t$ as 
\begin{equation}
Q=\int_{0}^{t}I(t')dt'.\label{eq:charge}
\end{equation}
The full counting statistics gives the cumulants of $Q$~\cite{levitov_electron_1996,jordan_transport_2004,sukhorukov_conditional_2007,singh_distribution_2016}, see also  Appendix \ref{sec:FCS}. The first and second cumulants of $Q$ are
\begin{equation}
\langle Q\rangle=t\langle I\rangle_{\Theta}=t[\langle I_{0}\rangle-f_{\Theta}(\varepsilon)\Delta I],\label{eq:Q-ave}
\end{equation}
\begin{equation}
\text{Var}(Q)=t\left[\frac{2(\Delta I)^{2}W_{\text{on}}W_{\text{off}}}{\Gamma^{3}}+\sum_{\alpha=0,\,1}S_{\alpha}P_{\alpha}\right].\label{eq:Q-var}
\end{equation}
Now we estimate the occupation $f$ by some specific sampling of $Q$ according to Eq.~(\ref{eq:Q-ave}). This results in the following estimator
\begin{equation}
\hat{f}=\frac{1}{\Delta I}\left[\frac{Q}{t}-\langle I_{0}\rangle\right].\label{eq:fhat}
\end{equation}
This estimator has a simple physical interpretation: the occupation of the dot is estimated from the duration of time spent on the dot, divided by the total time of the experiment. Note that in the long time limit, the distribution of $Q$ is approximately Gaussian. Then the occupation estimator (\ref{eq:fhat}) is also the maximum likelihood estimator. According to Eq.~(\ref{eq:Q-ave}), it is an unbiased estimator, i.e., $\langle\hat{f}\rangle=f_\Theta(\varepsilon)$. The variance of such an estimator is
\begin{align}
\text{Var}(\hat{f}) & =\frac{1}{t}\left[\frac{2}{\Gamma}f_{\Theta}(\varepsilon)[1-f_{\Theta}(\varepsilon)]+\sum_{\alpha=0,\,1}P_{\alpha}\tau_{\alpha}\right].
\end{align}
For sufficiently long $t$ such that $\text{Var}(\hat{f})$ is small, we can calculate the variance of the temperature estimator through the following error propagation relation \begin{equation}
\delta\hat{\Theta}=\left(\frac{df_\Theta(\varepsilon)}{d\Theta}\right)^{-1}\delta\hat{f}.\label{eq:err-propag}
\end{equation}
From Eq.~(\ref{eq:err-propag}), we find 
\begin{equation}
\text{Var}(\hat{\Theta})=\frac{2\Theta^{2}}{\xi(\varepsilon/\Theta)t}\left[\frac{1}{\Gamma}+\frac{\tau_{0}}{2f_\Theta(\varepsilon)}+\frac{\tau_{1}}{2[1-f_\Theta(\varepsilon)]}\right].\label{eq:full-var}
\end{equation}
Direct minimization of Eq.~(\ref{eq:full-var}) is tedious but can be done numerically. As in Sec.~\ref{subsec:Standard-protocol}, we assume the practical situation $\tau_{\alpha}\Gamma\ll1$ to simplify the minimization. With this assumption, we see that as long as the detuning $\big|\varepsilon\big|$ is not much larger than $\Theta$, such that $f_\Theta(\varepsilon)$ is neither close to $0$ nor $1$, the contributions from the shot noise in Eq.~(\ref{eq:full-var}) can be safely ignored. Therefore Eq.~(\ref{eq:full-var}) reduces to
\begin{equation}
\text{Var}(\hat{\Theta})=\frac{2\Theta^{2}}{\xi(\varepsilon/\Theta)\Gamma t},\label{eq:always-var}
\end{equation}
which is independent of the QPC parameters as in the standard protocol. When $\xi(x)$ is maximized at $\big|\varepsilon\big|/\Theta=2.4$, and consequently $f_\Theta(\varepsilon)\approx 10^{-2}$, the shot noise can be neglected as long as $\tau_{\alpha}\Gamma\ll10^{-2}$. In this case the minimum variance is
\begin{equation}
[\text{Var}(\hat{\Theta})]_{\min}=\frac{4.6\Theta^{2}}{\Gamma t}.\label{eq:always-min-var}
\end{equation}
Eq.~(\ref{eq:always-min-var}) is confirmed by Monte Carlo simulation as shown in Fig. \ref{fig:MC}, where we simulate the telegraph process to generate the measurement signals in the QPC. As in the standard protocol, the optimal regime requires the knowledge of the true value of $\Theta$ and therefore adaptive measurements are required in general. The behavior of the temperature estimator in a non-adaptive measurement is qualitatively the same as the standard protocol, as shown in Fig.~\ref{fig:sens-temp}. However, near the optimal regime the always-on protocol has a factor of 5 smaller variance than the standard protocol, as shown in Fig.~\ref{fig:MC}.

\section{\label{sec:conclusion}Conclusion}
We have analyzed a setup consisting of a quantum dot capacitively coupled to a QPC as shown in Fig.~\ref{fig:three-terminal-device} used as a nanoscale thermal transistor and noninvasive thermometer. The basic operation principle relies on the sensitivity of the average charge and heat current through the QPC to the average occupation of the quantum dot. The average dot occupation in turn depends on the temperature of the base reservoir. 
We characterized the performance of the thermal transistor by its power gain as well as the differential sensitivity of the average charge current through the QPC to a variation of the temperature of the base reservoir. The performance of the thermometer was characterized by the variance of the temperature estimator. Furthermore, the thermometer is noninvasive since  reading out the temperature only involves single electron tunneling and it is assumed that there is no energy exchange between the base and the QPC . 

Interestingly, as a consequence of the common operation principles, we have found that for both types of devices the maximal sensitivity occurs when the addition energy of the dot and the temperature of the base reservoir are related as $\varepsilon=2.4\Theta$. However, while for the transistor the base temperature is optimized at fixed addition energy, for the thermometer the optimization has to be performed over the addition energy keeping the base temperature fixed. Furthermore, while the sensitivity of the transistor depends on the difference $\Delta I$ of the average QPC currents, the sensitivity of the thermometer,  characterized by the variance of the temperature estimator, is independent of $\Delta I$.

The setup proposed here has already been implemented experimentally \cite{gasparinetti2012nongalvanic,torresani_nongalvanic_2013,mavalankar_non-invasive_2013,maradan_gaas_2014} for thermometers. Our work shows its interest for the purpose of controlling charge or heat flow at the nanoscale and sets its theoretical ideal performance.

\begin{acknowledgments}
We thank the KITP for hosting the program Thermodynamics of Quantum Systems:
Measurement, engines, and control, where this work was initiated. This research was supported in part by the National Science Foundation under Grant No. NSF PHY-1748958. JY, CE, and ANJ acknowledge the support from the U.S. Department of Energy, Office of Science, Office of Basic Energy Sciences under Award Number DE-SC-0017890 and the National Science Foundation under Grant No. NSF PHY-1748958. JS acknowledges funding from the Swedish VR and the Knut and Alice Wallenberg foundation through an Academy Fellowship.  BS acknowledges financial support from the Ministry of Innovation NRW via the “Programm zur F\"orderung der R\"uckkehr des hochqualifizierten Forschungsnachwuchses aus dem Ausland”. RS acknowledges support from the Ram\'on y Cajal program RYC-2016-20778 and the ``Mar\'ia de Maeztu\textquotedblright{} Programme for Units of Excellence in R\&D (MDM-2014-0377).
\end{acknowledgments}

\appendix

\section{\label{sec:FCS}Full counting statistics of the transmitted charge
in the QPC}

In this appendix we give a brief overview of the full counting statistics method applied to the current through the QPC, used in Sec.~\ref{sec_always_on}. We note that the statistical dependence of the transferred charges in consecutive time intervals, which are much larger than the typical correlation time of the current flowing in the QPC, can be neglected. Namely, for two consecutive time intervals $\delta t_{1}$, $\delta t_{2}$ where $\delta t_{1},\,\delta t_{2}\gg\Gamma^{-1}$, the charges transferred in the QPC denoted as $\delta Q_{1}$ and $\delta Q_{2}$ respectively can be treated as statistically independent. We define the total charge transferred during the two intervals as $\delta Q=\delta Q_{1}+\delta Q_{2}$ and denote the probability distributions of $\delta Q_{1}$, $\delta Q_{2}$ and $\delta Q$ as $P(\delta Q_{1},\,\delta t_{1})$, $P(\delta Q_{2},\,\delta t_{2})$ and $P(\delta Q,\,\delta t)$ respectively, where $\delta t=\delta t_{1}+\delta t_{2}$. Then $P(\delta Q,\,\delta t)$ is a convolution between $P(\delta Q_{1},\,\delta t_{1})$ and $P(\delta Q_2,\,\delta t_2)$ due to the statistical independence of $\delta Q_{1}$ and $\delta Q_{2}$ .

The moment and cumulant generating functions associated with $P(\delta Q,\,\delta t)$ are
\begin{equation}
G(\lambda,\,\delta t)=\int d(\delta Q)\exp(\text{i}\lambda\delta Q)P(\delta Q,\,\delta t),
\end{equation}
\begin{equation}
F(\lambda,\,\delta t)=\ln G(\lambda,\,\delta t).
\end{equation}
Thus we have 
\begin{equation}
G(\lambda,\,\delta t)=G(\lambda,\,\delta t_{1})G(\lambda,\,\delta t_{2}),
\end{equation}
\begin{equation}
F(\lambda,\,\delta t)=F(\lambda,\,\delta t_{1})+F(\lambda,\,\delta t_{2}).
\end{equation}
As a consequence, the cumulant generating function of $P(Q,\,t)$, where $Q$ is defined in Eq.~(\ref{eq:charge}) takes the form
\begin{equation}
F(\lambda,\,t)\equiv tH(\lambda).
\end{equation}
We Taylor expand $F(\lambda,\,t)$ as
\begin{equation}
F(\lambda,\,t)= \sum_{n=0}^{\infty}\frac{\lambda^{n}}{n!}\llangle Q^{n}\rrangle,
\end{equation}
and define
\begin{equation}
\llangle I^{n}\rrangle\equiv \llangle Q^{n}\rrangle/t.\label{eq:culmQ}
\end{equation}
Then, $H(\lambda,\,t)$ can be written as
\begin{equation}
H(\lambda)=\sum_{n=0}^{\infty}\frac{\lambda^{n}}{n!}\llangle I^{n}\rrangle.\label{eq:H(lamb)}
\end{equation}

We denote the probability $P_{\alpha}(Q,\,t)$ as the probability of transferring charge $Q$ up to time $t$ and having the dot in state $\alpha$ at time $t$. We note that if there is no tunneling between the dot and the base reservoir, 
\begin{align}
P_{\alpha}(Q,\,t) & =\int\dfrac{d\lambda}{2\pi}\exp[-i\lambda Q+F_{\alpha}(Q,\,t)]\nonumber \\
 & =\int\dfrac{d\lambda}{2\pi}\exp[-i\lambda Q+tH_{\alpha}(\lambda)],
\end{align}
which is equivalent to writing the time-derivative as 
\begin{equation}
\dot{P}_{\alpha}(Q,\,t)=\int\dfrac{d\lambda}{2\pi}\exp[-i\lambda Q]G_{\alpha}(\lambda,\,t)H_{\alpha}(\lambda).
\end{equation}
On top of this effect, we have to take into account the effect due to tunneling, which yields the master equation
\begin{align}
\dot{P}_{0}(Q,\,t) & =\int\dfrac{d\lambda}{2\pi}\exp[-i\lambda Q]G_{0}(\lambda,\,t)H_{0}(\lambda)\nonumber \\
 & -W_{\text{off}}P_{0}(Q,\,t)+W_{\text{on}}P_{1}(Q,\,t),
\end{align}
\begin{align}
\dot{P}_{1}(Q,\,t) & =\int\dfrac{d\lambda}{2\pi}\exp[-i\lambda Q]G_{1}(\lambda,\,t)H_{1}(\lambda)\nonumber \\
 & -W_{\text{on}}P_{1}(Q,\,t)+W_{\text{off}}P_{0}(Q,\,t).
\end{align}
Rewriting both sides in terms of the generation functions $G_{\alpha}(\lambda,\,t)=\exp[tH_{\alpha}(\lambda)]=\int dQ\exp(\text{i}\lambda Q)P_{\alpha}(Q,\,t)$ gives
\begin{align}
\dot{G}_{0}(\lambda,\,t) & =  [H_{0}(\lambda)-W_{\text{off}}]G_{0}(\lambda,t)+W_{\text{on}}G_{1}(\lambda,\,t),\\
\dot{G}_{1}(\lambda,\,t) & =  W_{\text{off}}G_{0}(\lambda,t)+[H_{1}(\lambda)-W_{\text{on}}]G_{1}(\lambda,t).
\end{align}
The above equation can be rewritten as 
\begin{equation}
\dot{\boldsymbol{G}}=\boldsymbol{H}\boldsymbol{G},\label{eq:G-diff}
\end{equation}
where 
\begin{equation}
\boldsymbol{G}(\lambda,\,t)\equiv\begin{bmatrix}G_{0}(\lambda,\,t)\\
G_{1}(\lambda,\,t)
\end{bmatrix},
\end{equation}
and
\begin{equation}
\boldsymbol{H}(\lambda)\equiv\begin{bmatrix}H_{0}(\lambda)-W_{\text{off}} & W_{\text{on}}\\
W_{\text{off}} & H_{1}(\lambda)-W_{\text{on}}
\end{bmatrix}.
\end{equation}
Thus the solution to Eq.~(\ref{eq:G-diff}) is
\begin{equation}
\boldsymbol{G}(\lambda,\,t)=\exp[t\boldsymbol{H}(\lambda)]\boldsymbol{G}(\lambda,\,0).
\end{equation}
When $t$ is sufficient large, the unconditional cumulant generating function 
\begin{equation}
H(\lambda)=\lim_{t\to\infty}\frac{\ln[\sum_{\alpha}G_{\alpha}(\lambda,\,t)]}{t}
\end{equation}
approaches the maximum eigenvalue of $\boldsymbol{H}(\lambda)$, which gives
\begin{align}
H(\lambda) & =\frac{1}{2}[\sum_{\alpha}H_{\alpha}(\lambda)-\Gamma]\nonumber \\
 & +\sqrt{[H_{1}(\lambda)-H_{0}(\lambda)-\Delta\Gamma]^{2}/4+W_{\text{on}}W_{\text{off}}}.
\end{align}
From this equation one can find 
\begin{equation}
\llangle I\rrangle=\partial H(\lambda)/\partial\lambda\big|_{\lambda=0}=\sum_{\alpha}P_{\alpha}\llangle I_{\alpha}\rrangle,\label{eq:Iculm}
\end{equation}
\begin{align}
\llangle I^{2}\rrangle & =\partial^{2}H(\lambda)/\partial\lambda^{2}\big|_{\lambda=0}\nonumber \\
 & =2(\Delta I)^{2}W_{\text{on}}W_{\text{off}}/\Gamma^{3}+\sum_{\alpha=0,\,1}\llangle I_{\alpha}^{2}\rrangle P_{\alpha\Theta},\label{eq:I2culm}
\end{align}
where $\llangle I_{\alpha}\rrangle=\partial H_{\alpha}(\lambda)/\partial\lambda\big|_{\lambda=0}$ and $\llangle I_{\alpha}^{2}\rrangle=\partial^{2}H_{\alpha}(\lambda)/\partial\lambda^{2}\big|_{\lambda=0}$ . In the long time limit, we can easily obtain
\begin{equation}
\llangle I_{\alpha}\rrangle=\lim_{t\to\infty}\frac{\llangle Q_\alpha \rrangle}{t}=\lim_{t\to\infty}\frac{1}{t}\int_{0}^{t}\langle I_\alpha(\tau)\rangle d\tau=\langle I_{\alpha}\rangle,\label{eq:ll-I-alpha-rr}
\end{equation}
and
\begin{align}
\llangle I_{\alpha}^{2}\rrangle= & \lim_{t\to\infty}\frac{\llangle Q_\alpha^{2}\rrangle}{t}\nonumber \\
= & \lim_{t\to\infty}\frac{1}{t}\left[\int_{0}^{t}\int_{0}^{t}\langle I_\alpha(\tau_{1})I_\alpha(\tau_{2})\rangle-\langle I_{\alpha}\rangle^{2}t^{2}\right]\nonumber \\
= & \lim_{t\to\infty}\int_{-t}^{t}\langle\delta I_\alpha(\tau)\delta I_\alpha(0)\rangle d\tau\nonumber \\
= & S_{\alpha},\label{eq:ll-Ialpha2-rr}
\end{align}
where $\delta I_\alpha(\tau)\equiv I_\alpha(\tau)-\langle I_{\alpha}\rangle$, $\langle I_{\alpha}\rangle$ and $S_{\alpha}$ are conditional average electric current in the QPC and the shot noise power spectral density defined in Eqs. (\ref{eq:Ialph-ave}, \ref{eq:Salpha}) respectively. With Eqs. (\ref{eq:ll-I-alpha-rr}, \ref{eq:ll-Ialpha2-rr}, \ref{eq:Iculm}, \ref{eq:I2culm}), one can easily obtain Eqs. (\ref{eq:Q-ave}, \ref{eq:Q-var}) in the main text.

\bibliography{Meine_Bibliothek.bib}

\begin{thebibliography}{95}%
\makeatletter
\providecommand \@ifxundefined [1]{%
 \@ifx{#1\undefined}
}%
\providecommand \@ifnum [1]{%
 \ifnum #1\expandafter \@firstoftwo
 \else \expandafter \@secondoftwo
 \fi
}%
\providecommand \@ifx [1]{%
 \ifx #1\expandafter \@firstoftwo
 \else \expandafter \@secondoftwo
 \fi
}%
\providecommand \natexlab [1]{#1}%
\providecommand \enquote  [1]{``#1''}%
\providecommand \bibnamefont  [1]{#1}%
\providecommand \bibfnamefont [1]{#1}%
\providecommand \citenamefont [1]{#1}%
\providecommand \href@noop [0]{\@secondoftwo}%
\providecommand \href [0]{\begingroup \@sanitize@url \@href}%
\providecommand \@href[1]{\@@startlink{#1}\@@href}%
\providecommand \@@href[1]{\endgroup#1\@@endlink}%
\providecommand \@sanitize@url [0]{\catcode `\\12\catcode `\$12\catcode
  `\&12\catcode `\#12\catcode `\^12\catcode `\_12\catcode `\%12\relax}%
\providecommand \@@startlink[1]{}%
\providecommand \@@endlink[0]{}%
\providecommand \url  [0]{\begingroup\@sanitize@url \@url }%
\providecommand \@url [1]{\endgroup\@href {#1}{\urlprefix }}%
\providecommand \urlprefix  [0]{URL }%
\providecommand \Eprint [0]{\href }%
\providecommand \doibase [0]{http://dx.doi.org/}%
\providecommand \selectlanguage [0]{\@gobble}%
\providecommand \bibinfo  [0]{\@secondoftwo}%
\providecommand \bibfield  [0]{\@secondoftwo}%
\providecommand \translation [1]{[#1]}%
\providecommand \BibitemOpen [0]{}%
\providecommand \bibitemStop [0]{}%
\providecommand \bibitemNoStop [0]{.\EOS\space}%
\providecommand \EOS [0]{\spacefactor3000\relax}%
\providecommand \BibitemShut  [1]{\csname bibitem#1\endcsname}%
\let\auto@bib@innerbib\@empty
\bibitem [{\citenamefont {Sothmann}\ \emph {et~al.}(2015)\citenamefont
  {Sothmann}, \citenamefont {S\'anchez},\ and\ \citenamefont
  {Jordan}}]{sothmann_thermoelectric_2015}%
  \BibitemOpen
  \bibfield  {author} {\bibinfo {author} {\bibfnamefont {B.}~\bibnamefont
  {Sothmann}}, \bibinfo {author} {\bibfnamefont {R.}~\bibnamefont {S\'anchez}},
  \ and\ \bibinfo {author} {\bibfnamefont {A.~N.}\ \bibnamefont {Jordan}},\
  }\href {\doibase 10.1088/0957-4484/26/3/032001} {\bibfield  {journal}
  {\bibinfo  {journal} {Nanotechnology}\ }\textbf {\bibinfo {volume} {26}},\
  \bibinfo {pages} {032001} (\bibinfo {year} {2015})}\BibitemShut {NoStop}%
\bibitem [{\citenamefont {Benenti}\ \emph {et~al.}(2017)\citenamefont
  {Benenti}, \citenamefont {Casati}, \citenamefont {Saito},\ and\ \citenamefont
  {Whitney}}]{benenti_fundamental_2017}%
  \BibitemOpen
  \bibfield  {author} {\bibinfo {author} {\bibfnamefont {G.}~\bibnamefont
  {Benenti}}, \bibinfo {author} {\bibfnamefont {G.}~\bibnamefont {Casati}},
  \bibinfo {author} {\bibfnamefont {K.}~\bibnamefont {Saito}}, \ and\ \bibinfo
  {author} {\bibfnamefont {R.}~\bibnamefont {Whitney}},\ }\href {\doibase
  10.1016/j.physrep.2017.05.008} {\bibfield  {journal} {\bibinfo  {journal}
  {Phys. Rep.}\ }\textbf {\bibinfo {volume} {694}},\ \bibinfo {pages} {1}
  (\bibinfo {year} {2017})}\BibitemShut {NoStop}%
\bibitem [{\citenamefont {Staring}\ \emph {et~al.}(1993)\citenamefont
  {Staring}, \citenamefont {Molenkamp}, \citenamefont {Alphenaar},
  \citenamefont {van Houten}, \citenamefont {Buyk}, \citenamefont {Mabesoone},
  \citenamefont {Beenakker},\ and\ \citenamefont
  {Foxon}}]{staring_coulomb-blockade_1993}%
  \BibitemOpen
  \bibfield  {author} {\bibinfo {author} {\bibfnamefont {A.~A.~M.}\
  \bibnamefont {Staring}}, \bibinfo {author} {\bibfnamefont {L.~W.}\
  \bibnamefont {Molenkamp}}, \bibinfo {author} {\bibfnamefont {B.~W.}\
  \bibnamefont {Alphenaar}}, \bibinfo {author} {\bibfnamefont {H.}~\bibnamefont
  {van Houten}}, \bibinfo {author} {\bibfnamefont {O.~J.~A.}\ \bibnamefont
  {Buyk}}, \bibinfo {author} {\bibfnamefont {M.~A.~A.}\ \bibnamefont
  {Mabesoone}}, \bibinfo {author} {\bibfnamefont {C.~W.~J.}\ \bibnamefont
  {Beenakker}}, \ and\ \bibinfo {author} {\bibfnamefont {C.~T.}\ \bibnamefont
  {Foxon}},\ }\href {\doibase 10.1209/0295-5075/22/1/011} {\bibfield  {journal}
  {\bibinfo  {journal} {Europhysics Letters (EPL)}\ }\textbf {\bibinfo {volume}
  {22}},\ \bibinfo {pages} {57} (\bibinfo {year} {1993})}\BibitemShut {NoStop}%
\bibitem [{\citenamefont {Dzurak}\ \emph {et~al.}(1993)\citenamefont {Dzurak},
  \citenamefont {Smith}, \citenamefont {Pepper}, \citenamefont {Ritchie},
  \citenamefont {Frost}, \citenamefont {Jones},\ and\ \citenamefont
  {Hasko}}]{dzurak_observation_1993}%
  \BibitemOpen
  \bibfield  {author} {\bibinfo {author} {\bibfnamefont {A.~S.}\ \bibnamefont
  {Dzurak}}, \bibinfo {author} {\bibfnamefont {C.~G.}\ \bibnamefont {Smith}},
  \bibinfo {author} {\bibfnamefont {M.}~\bibnamefont {Pepper}}, \bibinfo
  {author} {\bibfnamefont {D.~A.}\ \bibnamefont {Ritchie}}, \bibinfo {author}
  {\bibfnamefont {J.~E.~F.}\ \bibnamefont {Frost}}, \bibinfo {author}
  {\bibfnamefont {G.~A.~C.}\ \bibnamefont {Jones}}, \ and\ \bibinfo {author}
  {\bibfnamefont {D.~G.}\ \bibnamefont {Hasko}},\ }\href {\doibase
  10.1016/0038-1098(93)90819-9} {\bibfield  {journal} {\bibinfo  {journal}
  {Solid State Commun.}\ }\textbf {\bibinfo {volume} {87}},\ \bibinfo {pages}
  {1145} (\bibinfo {year} {1993})}\BibitemShut {NoStop}%
\bibitem [{\citenamefont {Humphrey}\ \emph {et~al.}(2002)\citenamefont
  {Humphrey}, \citenamefont {Newbury}, \citenamefont {Taylor},\ and\
  \citenamefont {Linke}}]{humphrey_reversible_2002}%
  \BibitemOpen
  \bibfield  {author} {\bibinfo {author} {\bibfnamefont {T.~E.}\ \bibnamefont
  {Humphrey}}, \bibinfo {author} {\bibfnamefont {R.}~\bibnamefont {Newbury}},
  \bibinfo {author} {\bibfnamefont {R.~P.}\ \bibnamefont {Taylor}}, \ and\
  \bibinfo {author} {\bibfnamefont {H.}~\bibnamefont {Linke}},\ }\href
  {\doibase 10.1103/PhysRevLett.89.116801} {\bibfield  {journal} {\bibinfo
  {journal} {Phys. Rev. Lett.}\ }\textbf {\bibinfo {volume} {89}},\ \bibinfo
  {pages} {116801} (\bibinfo {year} {2002})}\BibitemShut {NoStop}%
\bibitem [{\citenamefont {Entin-Wohlman}\ \emph {et~al.}(2010)\citenamefont
  {Entin-Wohlman}, \citenamefont {Imry},\ and\ \citenamefont
  {Aharony}}]{entin-wohlman_three-terminal_2010}%
  \BibitemOpen
  \bibfield  {author} {\bibinfo {author} {\bibfnamefont {O.}~\bibnamefont
  {Entin-Wohlman}}, \bibinfo {author} {\bibfnamefont {Y.}~\bibnamefont {Imry}},
  \ and\ \bibinfo {author} {\bibfnamefont {A.}~\bibnamefont {Aharony}},\ }\href
  {\doibase 10.1103/PhysRevB.82.115314} {\bibfield  {journal} {\bibinfo
  {journal} {Phys. Rev. B}\ }\textbf {\bibinfo {volume} {82}},\ \bibinfo
  {pages} {115314} (\bibinfo {year} {2010})}\BibitemShut {NoStop}%
\bibitem [{\citenamefont {S\'anchez}\ and\ \citenamefont
  {B\"uttiker}(2011)}]{sanchez_optimal_2011}%
  \BibitemOpen
  \bibfield  {author} {\bibinfo {author} {\bibfnamefont {R.}~\bibnamefont
  {S\'anchez}}\ and\ \bibinfo {author} {\bibfnamefont {M.}~\bibnamefont
  {B\"uttiker}},\ }\href {\doibase 10.1103/PhysRevB.83.085428} {\bibfield
  {journal} {\bibinfo  {journal} {Phys. Rev. B}\ }\textbf {\bibinfo {volume}
  {83}},\ \bibinfo {pages} {085428} (\bibinfo {year} {2011})}\BibitemShut
  {NoStop}%
\bibitem [{\citenamefont {Sothmann}\ \emph {et~al.}(2012)\citenamefont
  {Sothmann}, \citenamefont {S\'anchez}, \citenamefont {Jordan},\ and\
  \citenamefont {B\"uttiker}}]{sothmann_rectification_2012}%
  \BibitemOpen
  \bibfield  {author} {\bibinfo {author} {\bibfnamefont {B.}~\bibnamefont
  {Sothmann}}, \bibinfo {author} {\bibfnamefont {R.}~\bibnamefont {S\'anchez}},
  \bibinfo {author} {\bibfnamefont {A.~N.}\ \bibnamefont {Jordan}}, \ and\
  \bibinfo {author} {\bibfnamefont {M.}~\bibnamefont {B\"uttiker}},\ }\href
  {\doibase 10.1103/PhysRevB.85.205301} {\bibfield  {journal} {\bibinfo
  {journal} {Phys. Rev. B}\ }\textbf {\bibinfo {volume} {85}},\ \bibinfo
  {pages} {205301} (\bibinfo {year} {2012})}\BibitemShut {NoStop}%
\bibitem [{\citenamefont {Sothmann}\ and\ \citenamefont
  {B\"uttiker}(2012)}]{sothmann_magnon-driven_2012}%
  \BibitemOpen
  \bibfield  {author} {\bibinfo {author} {\bibfnamefont {B.}~\bibnamefont
  {Sothmann}}\ and\ \bibinfo {author} {\bibfnamefont {M.}~\bibnamefont
  {B\"uttiker}},\ }\href {\doibase 10.1209/0295-5075/99/27001} {\bibfield
  {journal} {\bibinfo  {journal} {Europhys. Lett.}\ }\textbf {\bibinfo {volume}
  {99}},\ \bibinfo {pages} {27001} (\bibinfo {year} {2012})}\BibitemShut
  {NoStop}%
\bibitem [{\citenamefont {Bergenfeldt}\ \emph {et~al.}(2014)\citenamefont
  {Bergenfeldt}, \citenamefont {Samuelsson}, \citenamefont {Sothmann},
  \citenamefont {Flindt},\ and\ \citenamefont
  {B\"uttiker}}]{bergenfeldt_hybrid_2014}%
  \BibitemOpen
  \bibfield  {author} {\bibinfo {author} {\bibfnamefont {C.}~\bibnamefont
  {Bergenfeldt}}, \bibinfo {author} {\bibfnamefont {P.}~\bibnamefont
  {Samuelsson}}, \bibinfo {author} {\bibfnamefont {B.}~\bibnamefont
  {Sothmann}}, \bibinfo {author} {\bibfnamefont {C.}~\bibnamefont {Flindt}}, \
  and\ \bibinfo {author} {\bibfnamefont {M.}~\bibnamefont {B\"uttiker}},\
  }\href {\doibase 10.1103/PhysRevLett.112.076803} {\bibfield  {journal}
  {\bibinfo  {journal} {Phys. Rev. Lett.}\ }\textbf {\bibinfo {volume} {112}},\
  \bibinfo {pages} {076803} (\bibinfo {year} {2014})}\BibitemShut {NoStop}%
\bibitem [{\citenamefont {S\'anchez}\ \emph
  {et~al.}(2015{\natexlab{a}})\citenamefont {S\'anchez}, \citenamefont
  {Sothmann},\ and\ \citenamefont {Jordan}}]{sanchez_chiral_2015}%
  \BibitemOpen
  \bibfield  {author} {\bibinfo {author} {\bibfnamefont {R.}~\bibnamefont
  {S\'anchez}}, \bibinfo {author} {\bibfnamefont {B.}~\bibnamefont {Sothmann}},
  \ and\ \bibinfo {author} {\bibfnamefont {A.~N.}\ \bibnamefont {Jordan}},\
  }\href {\doibase 10.1103/PhysRevLett.114.146801} {\bibfield  {journal}
  {\bibinfo  {journal} {Phys. Rev. Lett.}\ }\textbf {\bibinfo {volume} {114}},\
  \bibinfo {pages} {146801} (\bibinfo {year} {2015}{\natexlab{a}})}\BibitemShut
  {NoStop}%
\bibitem [{\citenamefont {Hofer}\ and\ \citenamefont
  {Sothmann}(2015)}]{hofer_quantum_2015}%
  \BibitemOpen
  \bibfield  {author} {\bibinfo {author} {\bibfnamefont {P.~P.}\ \bibnamefont
  {Hofer}}\ and\ \bibinfo {author} {\bibfnamefont {B.}~\bibnamefont
  {Sothmann}},\ }\href {\doibase 10.1103/PhysRevB.91.195406} {\bibfield
  {journal} {\bibinfo  {journal} {Phys. Rev. B}\ }\textbf {\bibinfo {volume}
  {91}},\ \bibinfo {pages} {195406} (\bibinfo {year} {2015})}\BibitemShut
  {NoStop}%
\bibitem [{\citenamefont {Roche}\ \emph {et~al.}(2015)\citenamefont {Roche},
  \citenamefont {Roulleau}, \citenamefont {Jullien}, \citenamefont {Jompol},
  \citenamefont {Farrer}, \citenamefont {Ritchie},\ and\ \citenamefont
  {Glattli}}]{roche_harvesting_2015}%
  \BibitemOpen
  \bibfield  {author} {\bibinfo {author} {\bibfnamefont {B.}~\bibnamefont
  {Roche}}, \bibinfo {author} {\bibfnamefont {P.}~\bibnamefont {Roulleau}},
  \bibinfo {author} {\bibfnamefont {T.}~\bibnamefont {Jullien}}, \bibinfo
  {author} {\bibfnamefont {Y.}~\bibnamefont {Jompol}}, \bibinfo {author}
  {\bibfnamefont {I.}~\bibnamefont {Farrer}}, \bibinfo {author} {\bibfnamefont
  {D.~A.}\ \bibnamefont {Ritchie}}, \ and\ \bibinfo {author} {\bibfnamefont
  {D.~C.}\ \bibnamefont {Glattli}},\ }\href {\doibase 10.1038/ncomms7738}
  {\bibfield  {journal} {\bibinfo  {journal} {Nat. Commun.}\ }\textbf {\bibinfo
  {volume} {6}} (\bibinfo {year} {2015}),\ 10.1038/ncomms7738}\BibitemShut
  {NoStop}%
\bibitem [{\citenamefont {Hartmann}\ \emph {et~al.}(2015)\citenamefont
  {Hartmann}, \citenamefont {Pfeffer}, \citenamefont
  {H{\ifmmode\ddot{o}\else\"{o}\fi}fling}, \citenamefont {Kamp},\ and\
  \citenamefont {Worschech}}]{hartmann_voltage_2015}%
  \BibitemOpen
  \bibfield  {author} {\bibinfo {author} {\bibfnamefont {F.}~\bibnamefont
  {Hartmann}}, \bibinfo {author} {\bibfnamefont {P.}~\bibnamefont {Pfeffer}},
  \bibinfo {author} {\bibfnamefont {S.}~\bibnamefont
  {H{\ifmmode\ddot{o}\else\"{o}\fi}fling}}, \bibinfo {author} {\bibfnamefont
  {M.}~\bibnamefont {Kamp}}, \ and\ \bibinfo {author} {\bibfnamefont
  {L.}~\bibnamefont {Worschech}},\ }\href {\doibase
  10.1103/PhysRevLett.114.146805} {\bibfield  {journal} {\bibinfo  {journal}
  {Phys. Rev. Lett.}\ }\textbf {\bibinfo {volume} {114}},\ \bibinfo {pages}
  {146805} (\bibinfo {year} {2015})}\BibitemShut {NoStop}%
\bibitem [{\citenamefont {Thierschmann}\ \emph
  {et~al.}(2015{\natexlab{a}})\citenamefont {Thierschmann}, \citenamefont
  {S\'anchez}, \citenamefont {Sothmann}, \citenamefont {Arnold}, \citenamefont
  {Heyn}, \citenamefont {Hansen}, \citenamefont {Buhmann},\ and\ \citenamefont
  {Molenkamp}}]{thierschmann_three-terminal_2015}%
  \BibitemOpen
  \bibfield  {author} {\bibinfo {author} {\bibfnamefont {H.}~\bibnamefont
  {Thierschmann}}, \bibinfo {author} {\bibfnamefont {R.}~\bibnamefont
  {S\'anchez}}, \bibinfo {author} {\bibfnamefont {B.}~\bibnamefont {Sothmann}},
  \bibinfo {author} {\bibfnamefont {F.}~\bibnamefont {Arnold}}, \bibinfo
  {author} {\bibfnamefont {C.}~\bibnamefont {Heyn}}, \bibinfo {author}
  {\bibfnamefont {W.}~\bibnamefont {Hansen}}, \bibinfo {author} {\bibfnamefont
  {H.}~\bibnamefont {Buhmann}}, \ and\ \bibinfo {author} {\bibfnamefont
  {L.~W.}\ \bibnamefont {Molenkamp}},\ }\href {\doibase 10.1038/nnano.2015.176}
  {\bibfield  {journal} {\bibinfo  {journal} {Nature Nanotech.}\ }\textbf
  {\bibinfo {volume} {10}},\ \bibinfo {pages} {854} (\bibinfo {year}
  {2015}{\natexlab{a}})}\BibitemShut {NoStop}%
\bibitem [{\citenamefont {Whitney}\ \emph {et~al.}(2016)\citenamefont
  {Whitney}, \citenamefont {S{\ifmmode\acute{a}\else\'{a}\fi}nchez},
  \citenamefont {Haupt},\ and\ \citenamefont
  {Splettstoesser}}]{Whitney2016Jan}%
  \BibitemOpen
  \bibfield  {author} {\bibinfo {author} {\bibfnamefont {R.~S.}\ \bibnamefont
  {Whitney}}, \bibinfo {author} {\bibfnamefont {R.}~\bibnamefont
  {S{\ifmmode\acute{a}\else\'{a}\fi}nchez}}, \bibinfo {author} {\bibfnamefont
  {F.}~\bibnamefont {Haupt}}, \ and\ \bibinfo {author} {\bibfnamefont
  {J.}~\bibnamefont {Splettstoesser}},\ }\href {\doibase
  10.1016/j.physe.2015.09.025} {\bibfield  {journal} {\bibinfo  {journal}
  {Physica E}\ }\textbf {\bibinfo {volume} {75}},\ \bibinfo {pages} {257}
  (\bibinfo {year} {2016})}\BibitemShut {NoStop}%
\bibitem [{\citenamefont {Schulenborg}\ \emph {et~al.}(2017)\citenamefont
  {Schulenborg}, \citenamefont {Di~Marco}, \citenamefont {Vanherck},
  \citenamefont {Wegewijs},\ and\ \citenamefont
  {Splettstoesser}}]{Schulenborg2017Dec}%
  \BibitemOpen
  \bibfield  {author} {\bibinfo {author} {\bibfnamefont {J.}~\bibnamefont
  {Schulenborg}}, \bibinfo {author} {\bibfnamefont {A.}~\bibnamefont
  {Di~Marco}}, \bibinfo {author} {\bibfnamefont {J.}~\bibnamefont {Vanherck}},
  \bibinfo {author} {\bibfnamefont {M.~R.}\ \bibnamefont {Wegewijs}}, \ and\
  \bibinfo {author} {\bibfnamefont {J.}~\bibnamefont {Splettstoesser}},\ }\href
  {\doibase 10.3390/e19120668} {\bibfield  {journal} {\bibinfo  {journal}
  {Entropy}\ }\textbf {\bibinfo {volume} {19}},\ \bibinfo {pages} {668}
  (\bibinfo {year} {2017})}\BibitemShut {NoStop}%
\bibitem [{\citenamefont {Josefsson}\ \emph {et~al.}(2018)\citenamefont
  {Josefsson}, \citenamefont {Svilans}, \citenamefont {Burke}, \citenamefont
  {Hoffmann}, \citenamefont {Fahlvik}, \citenamefont {Thelander}, \citenamefont
  {Leijnse},\ and\ \citenamefont {Linke}}]{josefsson_quantum-dot_2018}%
  \BibitemOpen
  \bibfield  {author} {\bibinfo {author} {\bibfnamefont {M.}~\bibnamefont
  {Josefsson}}, \bibinfo {author} {\bibfnamefont {A.}~\bibnamefont {Svilans}},
  \bibinfo {author} {\bibfnamefont {A.~M.}\ \bibnamefont {Burke}}, \bibinfo
  {author} {\bibfnamefont {E.~A.}\ \bibnamefont {Hoffmann}}, \bibinfo {author}
  {\bibfnamefont {S.}~\bibnamefont {Fahlvik}}, \bibinfo {author} {\bibfnamefont
  {C.}~\bibnamefont {Thelander}}, \bibinfo {author} {\bibfnamefont
  {M.}~\bibnamefont {Leijnse}}, \ and\ \bibinfo {author} {\bibfnamefont
  {H.}~\bibnamefont {Linke}},\ }\href {\doibase 10.1038/s41565-018-0200-5}
  {\bibfield  {journal} {\bibinfo  {journal} {Nat. Nanotechnol.}\ }\textbf
  {\bibinfo {volume} {13}},\ \bibinfo {pages} {920} (\bibinfo {year}
  {2018})}\BibitemShut {NoStop}%
\bibitem [{\citenamefont {S{\ifmmode\acute{a}\else\'{a}\fi}nchez}\ \emph
  {et~al.}(2018)\citenamefont {S{\ifmmode\acute{a}\else\'{a}\fi}nchez},
  \citenamefont {Splettstoesser},\ and\ \citenamefont
  {Whitney}}]{Sanchez2018Nov}%
  \BibitemOpen
  \bibfield  {author} {\bibinfo {author} {\bibfnamefont {R.}~\bibnamefont
  {S{\ifmmode\acute{a}\else\'{a}\fi}nchez}}, \bibinfo {author} {\bibfnamefont
  {J.}~\bibnamefont {Splettstoesser}}, \ and\ \bibinfo {author} {\bibfnamefont
  {R.~S.}\ \bibnamefont {Whitney}},\ }\href {https://arxiv.org/abs/1811.02453}
  {\bibfield  {journal} {\bibinfo  {journal} {arXiv}\ } (\bibinfo {year}
  {2018})},\ \Eprint {http://arxiv.org/abs/1811.02453} {1811.02453}
  \BibitemShut {NoStop}%
\bibitem [{\citenamefont {Giazotto}\ \emph {et~al.}(2006)\citenamefont
  {Giazotto}, \citenamefont {Heikkil\"a}, \citenamefont {Luukanen},
  \citenamefont {Savin},\ and\ \citenamefont
  {Pekola}}]{giazotto_opportunities_2006}%
  \BibitemOpen
  \bibfield  {author} {\bibinfo {author} {\bibfnamefont {F.}~\bibnamefont
  {Giazotto}}, \bibinfo {author} {\bibfnamefont {T.~T.}\ \bibnamefont
  {Heikkil\"a}}, \bibinfo {author} {\bibfnamefont {A.}~\bibnamefont
  {Luukanen}}, \bibinfo {author} {\bibfnamefont {A.~M.}\ \bibnamefont {Savin}},
  \ and\ \bibinfo {author} {\bibfnamefont {J.~P.}\ \bibnamefont {Pekola}},\
  }\href {\doibase 10.1103/RevModPhys.78.217} {\bibfield  {journal} {\bibinfo
  {journal} {Rev. Mod. Phys.}\ }\textbf {\bibinfo {volume} {78}},\ \bibinfo
  {pages} {217} (\bibinfo {year} {2006})}\BibitemShut {NoStop}%
\bibitem [{\citenamefont {Pekola}\ and\ \citenamefont
  {Hekking}(2007)}]{pekola_normal-metal-superconductor_2007}%
  \BibitemOpen
  \bibfield  {author} {\bibinfo {author} {\bibfnamefont {J.~P.}\ \bibnamefont
  {Pekola}}\ and\ \bibinfo {author} {\bibfnamefont {F.~W.~J.}\ \bibnamefont
  {Hekking}},\ }\href {\doibase 10.1103/PhysRevLett.98.210604} {\bibfield
  {journal} {\bibinfo  {journal} {Phys. Rev. Lett.}\ }\textbf {\bibinfo
  {volume} {98}},\ \bibinfo {pages} {210604} (\bibinfo {year}
  {2007})}\BibitemShut {NoStop}%
\bibitem [{\citenamefont {Edwards}\ \emph {et~al.}(1993)\citenamefont
  {Edwards}, \citenamefont {Niu},\ and\ \citenamefont
  {de~Lozanne}}]{edwards_quantum-dot_1993}%
  \BibitemOpen
  \bibfield  {author} {\bibinfo {author} {\bibfnamefont {H.~L.}\ \bibnamefont
  {Edwards}}, \bibinfo {author} {\bibfnamefont {Q.}~\bibnamefont {Niu}}, \ and\
  \bibinfo {author} {\bibfnamefont {A.~L.}\ \bibnamefont {de~Lozanne}},\ }\href
  {\doibase doi:10.1063/1.110672} {\bibfield  {journal} {\bibinfo  {journal}
  {Appl. Phys. Lett.}\ }\textbf {\bibinfo {volume} {63}},\ \bibinfo {pages}
  {1815} (\bibinfo {year} {1993})}\BibitemShut {NoStop}%
\bibitem [{\citenamefont {Prance}\ \emph {et~al.}(2009)\citenamefont {Prance},
  \citenamefont {Smith}, \citenamefont {Griffiths}, \citenamefont {Chorley},
  \citenamefont {Anderson}, \citenamefont {Jones}, \citenamefont {Farrer},\
  and\ \citenamefont {Ritchie}}]{prance_electronic_2009}%
  \BibitemOpen
  \bibfield  {author} {\bibinfo {author} {\bibfnamefont {J.~R.}\ \bibnamefont
  {Prance}}, \bibinfo {author} {\bibfnamefont {C.~G.}\ \bibnamefont {Smith}},
  \bibinfo {author} {\bibfnamefont {J.~P.}\ \bibnamefont {Griffiths}}, \bibinfo
  {author} {\bibfnamefont {S.~J.}\ \bibnamefont {Chorley}}, \bibinfo {author}
  {\bibfnamefont {D.}~\bibnamefont {Anderson}}, \bibinfo {author}
  {\bibfnamefont {G.~A.~C.}\ \bibnamefont {Jones}}, \bibinfo {author}
  {\bibfnamefont {I.}~\bibnamefont {Farrer}}, \ and\ \bibinfo {author}
  {\bibfnamefont {D.~A.}\ \bibnamefont {Ritchie}},\ }\href {\doibase
  10.1103/PhysRevLett.102.146602} {\bibfield  {journal} {\bibinfo  {journal}
  {Phys. Rev. Lett.}\ }\textbf {\bibinfo {volume} {102}},\ \bibinfo {pages}
  {146602} (\bibinfo {year} {2009})}\BibitemShut {NoStop}%
\bibitem [{\citenamefont {Zhang}\ \emph {et~al.}(2015)\citenamefont {Zhang},
  \citenamefont {Lin},\ and\ \citenamefont {Chen}}]{zhang_three-terminal_2015}%
  \BibitemOpen
  \bibfield  {author} {\bibinfo {author} {\bibfnamefont {Y.}~\bibnamefont
  {Zhang}}, \bibinfo {author} {\bibfnamefont {G.}~\bibnamefont {Lin}}, \ and\
  \bibinfo {author} {\bibfnamefont {J.}~\bibnamefont {Chen}},\ }\href {\doibase
  10.1103/PhysRevE.91.052118} {\bibfield  {journal} {\bibinfo  {journal} {Phys.
  Rev. E}\ }\textbf {\bibinfo {volume} {91}},\ \bibinfo {pages} {052118}
  (\bibinfo {year} {2015})}\BibitemShut {NoStop}%
\bibitem [{\citenamefont {Koski}\ \emph {et~al.}(2015)\citenamefont {Koski},
  \citenamefont {Kutvonen}, \citenamefont {Khaymovich}, \citenamefont
  {Ala-Nissila},\ and\ \citenamefont {Pekola}}]{koski_-chip_2015}%
  \BibitemOpen
  \bibfield  {author} {\bibinfo {author} {\bibfnamefont {J.}~\bibnamefont
  {Koski}}, \bibinfo {author} {\bibfnamefont {A.}~\bibnamefont {Kutvonen}},
  \bibinfo {author} {\bibfnamefont {I.}~\bibnamefont {Khaymovich}}, \bibinfo
  {author} {\bibfnamefont {T.}~\bibnamefont {Ala-Nissila}}, \ and\ \bibinfo
  {author} {\bibfnamefont {J.}~\bibnamefont {Pekola}},\ }\href {\doibase
  10.1103/PhysRevLett.115.260602} {\bibfield  {journal} {\bibinfo  {journal}
  {Phys. Rev. Lett.}\ }\textbf {\bibinfo {volume} {115}},\ \bibinfo {pages}
  {260602} (\bibinfo {year} {2015})}\BibitemShut {NoStop}%
\bibitem [{\citenamefont {Hofer}\ \emph {et~al.}(2016)\citenamefont {Hofer},
  \citenamefont {Perarnau-Llobet}, \citenamefont {Brask}, \citenamefont
  {Silva}, \citenamefont {Huber},\ and\ \citenamefont
  {Brunner}}]{hofer_autonomous_2016}%
  \BibitemOpen
  \bibfield  {author} {\bibinfo {author} {\bibfnamefont {P.~P.}\ \bibnamefont
  {Hofer}}, \bibinfo {author} {\bibfnamefont {M.}~\bibnamefont
  {Perarnau-Llobet}}, \bibinfo {author} {\bibfnamefont {J.~B.}\ \bibnamefont
  {Brask}}, \bibinfo {author} {\bibfnamefont {R.}~\bibnamefont {Silva}},
  \bibinfo {author} {\bibfnamefont {M.}~\bibnamefont {Huber}}, \ and\ \bibinfo
  {author} {\bibfnamefont {N.}~\bibnamefont {Brunner}},\ }\href {\doibase
  10.1103/PhysRevB.94.235420} {\bibfield  {journal} {\bibinfo  {journal} {Phys.
  Rev. B}\ }\textbf {\bibinfo {volume} {94}},\ \bibinfo {pages} {235420}
  (\bibinfo {year} {2016})}\BibitemShut {NoStop}%
\bibitem [{\citenamefont {S\'anchez}(2017)}]{sanchez_correlation-induced_2017}%
  \BibitemOpen
  \bibfield  {author} {\bibinfo {author} {\bibfnamefont {R.}~\bibnamefont
  {S\'anchez}},\ }\href {\doibase 10.1063/1.5008481} {\bibfield  {journal}
  {\bibinfo  {journal} {Appl. Phys. Lett.}\ }\textbf {\bibinfo {volume}
  {111}},\ \bibinfo {pages} {223103} (\bibinfo {year} {2017})}\BibitemShut
  {NoStop}%
\bibitem [{\citenamefont {Scheibner}\ \emph {et~al.}(2008)\citenamefont
  {Scheibner}, \citenamefont {K\"onig}, \citenamefont {Reuter}, \citenamefont
  {Wieck}, \citenamefont {Gould}, \citenamefont {Buhmann},\ and\ \citenamefont
  {Molenkamp}}]{scheibner_quantum_2008}%
  \BibitemOpen
  \bibfield  {author} {\bibinfo {author} {\bibfnamefont {R.}~\bibnamefont
  {Scheibner}}, \bibinfo {author} {\bibfnamefont {M.}~\bibnamefont {K\"onig}},
  \bibinfo {author} {\bibfnamefont {D.}~\bibnamefont {Reuter}}, \bibinfo
  {author} {\bibfnamefont {A.~D.}\ \bibnamefont {Wieck}}, \bibinfo {author}
  {\bibfnamefont {C.}~\bibnamefont {Gould}}, \bibinfo {author} {\bibfnamefont
  {H.}~\bibnamefont {Buhmann}}, \ and\ \bibinfo {author} {\bibfnamefont
  {L.~W.}\ \bibnamefont {Molenkamp}},\ }\href {\doibase
  10.1088/1367-2630/10/8/083016} {\bibfield  {journal} {\bibinfo  {journal}
  {New J. Phys.}\ }\textbf {\bibinfo {volume} {10}},\ \bibinfo {pages} {083016}
  (\bibinfo {year} {2008})}\BibitemShut {NoStop}%
\bibitem [{\citenamefont {Ruokola}\ and\ \citenamefont
  {Ojanen}(2011)}]{ruokola_single-electron_2011}%
  \BibitemOpen
  \bibfield  {author} {\bibinfo {author} {\bibfnamefont {T.}~\bibnamefont
  {Ruokola}}\ and\ \bibinfo {author} {\bibfnamefont {T.}~\bibnamefont
  {Ojanen}},\ }\href {\doibase 10.1103/PhysRevB.83.241404} {\bibfield
  {journal} {\bibinfo  {journal} {Phys. Rev. B}\ }\textbf {\bibinfo {volume}
  {83}},\ \bibinfo {pages} {241404} (\bibinfo {year} {2011})}\BibitemShut
  {NoStop}%
\bibitem [{\citenamefont {Fornieri}\ \emph {et~al.}(2014)\citenamefont
  {Fornieri}, \citenamefont {Mart\'inez-P\'erez},\ and\ \citenamefont
  {Giazotto}}]{fornieri_normal_2014}%
  \BibitemOpen
  \bibfield  {author} {\bibinfo {author} {\bibfnamefont {A.}~\bibnamefont
  {Fornieri}}, \bibinfo {author} {\bibfnamefont {M.~J.}\ \bibnamefont
  {Mart\'inez-P\'erez}}, \ and\ \bibinfo {author} {\bibfnamefont
  {F.}~\bibnamefont {Giazotto}},\ }\href {\doibase 10.1063/1.4875917}
  {\bibfield  {journal} {\bibinfo  {journal} {Appl. Phys. Lett.}\ }\textbf
  {\bibinfo {volume} {104}},\ \bibinfo {pages} {183108} (\bibinfo {year}
  {2014})}\BibitemShut {NoStop}%
\bibitem [{\citenamefont {Jiang}\ \emph {et~al.}(2015)\citenamefont {Jiang},
  \citenamefont {Kulkarni}, \citenamefont {Segal},\ and\ \citenamefont
  {Imry}}]{jiang_phonon_2015}%
  \BibitemOpen
  \bibfield  {author} {\bibinfo {author} {\bibfnamefont {J.-H.}\ \bibnamefont
  {Jiang}}, \bibinfo {author} {\bibfnamefont {M.}~\bibnamefont {Kulkarni}},
  \bibinfo {author} {\bibfnamefont {D.}~\bibnamefont {Segal}}, \ and\ \bibinfo
  {author} {\bibfnamefont {Y.}~\bibnamefont {Imry}},\ }\href {\doibase
  10.1103/PhysRevB.92.045309} {\bibfield  {journal} {\bibinfo  {journal} {Phys.
  Rev. B}\ }\textbf {\bibinfo {volume} {92}},\ \bibinfo {pages} {045309}
  (\bibinfo {year} {2015})}\BibitemShut {NoStop}%
\bibitem [{\citenamefont {S\'anchez}\ \emph
  {et~al.}(2015{\natexlab{b}})\citenamefont {S\'anchez}, \citenamefont
  {Sothmann},\ and\ \citenamefont {Jordan}}]{sanchez_heat_2015}%
  \BibitemOpen
  \bibfield  {author} {\bibinfo {author} {\bibfnamefont {R.}~\bibnamefont
  {S\'anchez}}, \bibinfo {author} {\bibfnamefont {B.}~\bibnamefont {Sothmann}},
  \ and\ \bibinfo {author} {\bibfnamefont {A.~N.}\ \bibnamefont {Jordan}},\
  }\href {\doibase 10.1088/1367-2630/17/7/075006} {\bibfield  {journal}
  {\bibinfo  {journal} {New J. Phys.}\ }\textbf {\bibinfo {volume} {17}},\
  \bibinfo {pages} {075006} (\bibinfo {year} {2015}{\natexlab{b}})}\BibitemShut
  {NoStop}%
\bibitem [{\citenamefont {Mart\'inez-P\'erez}\ \emph
  {et~al.}(2015)\citenamefont {Mart\'inez-P\'erez}, \citenamefont {Fornieri},\
  and\ \citenamefont {Giazotto}}]{martinez-perez_rectification_2015}%
  \BibitemOpen
  \bibfield  {author} {\bibinfo {author} {\bibfnamefont {M.~J.}\ \bibnamefont
  {Mart\'inez-P\'erez}}, \bibinfo {author} {\bibfnamefont {A.}~\bibnamefont
  {Fornieri}}, \ and\ \bibinfo {author} {\bibfnamefont {F.}~\bibnamefont
  {Giazotto}},\ }\href {\doibase 10.1038/nnano.2015.11} {\bibfield  {journal}
  {\bibinfo  {journal} {Nature Nanotech.}\ }\textbf {\bibinfo {volume} {10}},\
  \bibinfo {pages} {303} (\bibinfo {year} {2015})}\BibitemShut {NoStop}%
\bibitem [{\citenamefont {Li}\ \emph {et~al.}(2006)\citenamefont {Li},
  \citenamefont {Wang},\ and\ \citenamefont {Casati}}]{li_negative_2006}%
  \BibitemOpen
  \bibfield  {author} {\bibinfo {author} {\bibfnamefont {B.}~\bibnamefont
  {Li}}, \bibinfo {author} {\bibfnamefont {L.}~\bibnamefont {Wang}}, \ and\
  \bibinfo {author} {\bibfnamefont {G.}~\bibnamefont {Casati}},\ }\href
  {\doibase 10.1063/1.2191730} {\bibfield  {journal} {\bibinfo  {journal}
  {Appl. Phys. Lett.}\ }\textbf {\bibinfo {volume} {88}},\ \bibinfo {pages}
  {143501} (\bibinfo {year} {2006})}\BibitemShut {NoStop}%
\bibitem [{\citenamefont {Joulain}\ \emph {et~al.}(2016)\citenamefont
  {Joulain}, \citenamefont {Drevillon}, \citenamefont {Ezzahri},\ and\
  \citenamefont {Ordonez-Miranda}}]{joulain_quantum_2016}%
  \BibitemOpen
  \bibfield  {author} {\bibinfo {author} {\bibfnamefont {K.}~\bibnamefont
  {Joulain}}, \bibinfo {author} {\bibfnamefont {J.}~\bibnamefont {Drevillon}},
  \bibinfo {author} {\bibfnamefont {Y.}~\bibnamefont {Ezzahri}}, \ and\
  \bibinfo {author} {\bibfnamefont {J.}~\bibnamefont {Ordonez-Miranda}},\
  }\href {\doibase 10.1103/PhysRevLett.116.200601} {\bibfield  {journal}
  {\bibinfo  {journal} {Phys. Rev. Lett.}\ }\textbf {\bibinfo {volume} {116}},\
  \bibinfo {pages} {200601} (\bibinfo {year} {2016})}\BibitemShut {NoStop}%
\bibitem [{\citenamefont {S\'anchez}\ \emph
  {et~al.}(2017{\natexlab{a}})\citenamefont {S\'anchez}, \citenamefont
  {Thierschmann},\ and\ \citenamefont
  {Molenkamp}}]{sanchez_single-electron_2017}%
  \BibitemOpen
  \bibfield  {author} {\bibinfo {author} {\bibfnamefont {R.}~\bibnamefont
  {S\'anchez}}, \bibinfo {author} {\bibfnamefont {H.}~\bibnamefont
  {Thierschmann}}, \ and\ \bibinfo {author} {\bibfnamefont {L.~W.}\
  \bibnamefont {Molenkamp}},\ }\href {\doibase 10.1088/1367-2630/aa8b94}
  {\bibfield  {journal} {\bibinfo  {journal} {New J. Phys.}\ }\textbf {\bibinfo
  {volume} {19}},\ \bibinfo {pages} {113040} (\bibinfo {year}
  {2017}{\natexlab{a}})}\BibitemShut {NoStop}%
\bibitem [{\citenamefont {S\'anchez}\ \emph
  {et~al.}(2017{\natexlab{b}})\citenamefont {S\'anchez}, \citenamefont
  {Thierschmann},\ and\ \citenamefont {Molenkamp}}]{sanchez_all-thermal_2017}%
  \BibitemOpen
  \bibfield  {author} {\bibinfo {author} {\bibfnamefont {R.}~\bibnamefont
  {S\'anchez}}, \bibinfo {author} {\bibfnamefont {H.}~\bibnamefont
  {Thierschmann}}, \ and\ \bibinfo {author} {\bibfnamefont {L.~W.}\
  \bibnamefont {Molenkamp}},\ }\href {\doibase 10.1103/PhysRevB.95.241401}
  {\bibfield  {journal} {\bibinfo  {journal} {Phys. Rev. B}\ }\textbf {\bibinfo
  {volume} {95}},\ \bibinfo {pages} {241401} (\bibinfo {year}
  {2017}{\natexlab{b}})}\BibitemShut {NoStop}%
\bibitem [{\citenamefont {Zhang}\ \emph {et~al.}(2018)\citenamefont {Zhang},
  \citenamefont {Yang}, \citenamefont {Zhang}, \citenamefont {Lin},
  \citenamefont {Lin},\ and\ \citenamefont
  {Chen}}]{zhang_coulomb-coupled_2018}%
  \BibitemOpen
  \bibfield  {author} {\bibinfo {author} {\bibfnamefont {Y.}~\bibnamefont
  {Zhang}}, \bibinfo {author} {\bibfnamefont {Z.}~\bibnamefont {Yang}},
  \bibinfo {author} {\bibfnamefont {X.}~\bibnamefont {Zhang}}, \bibinfo
  {author} {\bibfnamefont {B.}~\bibnamefont {Lin}}, \bibinfo {author}
  {\bibfnamefont {G.}~\bibnamefont {Lin}}, \ and\ \bibinfo {author}
  {\bibfnamefont {J.}~\bibnamefont {Chen}},\ }\href {\doibase
  10.1209/0295-5075/122/17002} {\bibfield  {journal} {\bibinfo  {journal}
  {Europhys. Lett.}\ }\textbf {\bibinfo {volume} {122}},\ \bibinfo {pages}
  {17002} (\bibinfo {year} {2018})}\BibitemShut {NoStop}%
\bibitem [{\citenamefont {Tang}\ \emph {et~al.}(2019)\citenamefont {Tang},
  \citenamefont {Peng},\ and\ \citenamefont {Wang}}]{Tang2019Feb}%
  \BibitemOpen
  \bibfield  {author} {\bibinfo {author} {\bibfnamefont {G.}~\bibnamefont
  {Tang}}, \bibinfo {author} {\bibfnamefont {J.}~\bibnamefont {Peng}}, \ and\
  \bibinfo {author} {\bibfnamefont {J.-S.}\ \bibnamefont {Wang}},\ }\href
  {\doibase 10.1140/epjb/e2019-90747-0} {\bibfield  {journal} {\bibinfo
  {journal} {Eur. Phys. J. B}\ }\textbf {\bibinfo {volume} {92}},\ \bibinfo
  {pages} {27} (\bibinfo {year} {2019})}\BibitemShut {NoStop}%
\bibitem [{\citenamefont {Guo}\ \emph {et~al.}(2018)\citenamefont {Guo},
  \citenamefont {Liu},\ and\ \citenamefont {Yu}}]{guo2018quantum}%
  \BibitemOpen
  \bibfield  {author} {\bibinfo {author} {\bibfnamefont {B.-q.}\ \bibnamefont
  {Guo}}, \bibinfo {author} {\bibfnamefont {T.}~\bibnamefont {Liu}}, \ and\
  \bibinfo {author} {\bibfnamefont {C.-s.}\ \bibnamefont {Yu}},\ }\href
  {\doibase 10.1103/PhysRevE.98.022118} {\bibfield  {journal} {\bibinfo
  {journal} {Phys. Rev. E}\ }\textbf {\bibinfo {volume} {98}},\ \bibinfo
  {pages} {022118} (\bibinfo {year} {2018})}\BibitemShut {NoStop}%
\bibitem [{\citenamefont {Correa}\ \emph {et~al.}(2015)\citenamefont {Correa},
  \citenamefont {Mehboudi}, \citenamefont {Adesso},\ and\ \citenamefont
  {Sanpera}}]{correa_individual_2015}%
  \BibitemOpen
  \bibfield  {author} {\bibinfo {author} {\bibfnamefont {L.~A.}\ \bibnamefont
  {Correa}}, \bibinfo {author} {\bibfnamefont {M.}~\bibnamefont {Mehboudi}},
  \bibinfo {author} {\bibfnamefont {G.}~\bibnamefont {Adesso}}, \ and\ \bibinfo
  {author} {\bibfnamefont {A.}~\bibnamefont {Sanpera}},\ }\href {\doibase
  10.1103/PhysRevLett.114.220405} {\bibfield  {journal} {\bibinfo  {journal}
  {Phys. Rev. Lett.}\ }\textbf {\bibinfo {volume} {114}},\ \bibinfo {pages}
  {220405} (\bibinfo {year} {2015})}\BibitemShut {NoStop}%
\bibitem [{\citenamefont {Hofer}\ \emph {et~al.}(2017)\citenamefont {Hofer},
  \citenamefont {Brask}, \citenamefont {Perarnau-Llobet},\ and\ \citenamefont
  {Brunner}}]{hofer_quantum_2017}%
  \BibitemOpen
  \bibfield  {author} {\bibinfo {author} {\bibfnamefont {P.~P.}\ \bibnamefont
  {Hofer}}, \bibinfo {author} {\bibfnamefont {J.~B.}\ \bibnamefont {Brask}},
  \bibinfo {author} {\bibfnamefont {M.}~\bibnamefont {Perarnau-Llobet}}, \ and\
  \bibinfo {author} {\bibfnamefont {N.}~\bibnamefont {Brunner}},\ }\href
  {\doibase 10.1103/PhysRevLett.119.090603} {\bibfield  {journal} {\bibinfo
  {journal} {Phys. Rev. Lett.}\ }\textbf {\bibinfo {volume} {119}},\ \bibinfo
  {pages} {090603} (\bibinfo {year} {2017})}\BibitemShut {NoStop}%
\bibitem [{\citenamefont {Mehboudi}\ \emph {et~al.}(2018)\citenamefont
  {Mehboudi}, \citenamefont {Sanpera},\ and\ \citenamefont
  {Correa}}]{mehboudi_thermometry_2018}%
  \BibitemOpen
  \bibfield  {author} {\bibinfo {author} {\bibfnamefont {M.}~\bibnamefont
  {Mehboudi}}, \bibinfo {author} {\bibfnamefont {A.}~\bibnamefont {Sanpera}}, \
  and\ \bibinfo {author} {\bibfnamefont {L.~A.}\ \bibnamefont {Correa}},\
  }\href {http://arxiv.org/abs/1811.03988} {\bibfield  {journal} {\bibinfo
  {journal} {arXiv:1811.03988}\ } (\bibinfo {year} {2018})},\ \bibinfo {note}
  {arXiv: 1811.03988}\BibitemShut {NoStop}%
\bibitem [{\citenamefont {De~Pasquale}\ and\ \citenamefont
  {Stace}(2018)}]{de_pasquale_quantum_2018}%
  \BibitemOpen
  \bibfield  {author} {\bibinfo {author} {\bibfnamefont {A.}~\bibnamefont
  {De~Pasquale}}\ and\ \bibinfo {author} {\bibfnamefont {T.~M.}\ \bibnamefont
  {Stace}},\ }\href {http://arxiv.org/abs/1807.05762} {\bibfield  {journal}
  {\bibinfo  {journal} {arXiv:1807.05762}\ } (\bibinfo {year} {2018})},\
  \bibinfo {note} {arXiv: 1807.05762}\BibitemShut {NoStop}%
\bibitem [{\citenamefont {Spietz}\ \emph {et~al.}(2003)\citenamefont {Spietz},
  \citenamefont {Lehnert}, \citenamefont {Siddiqi},\ and\ \citenamefont
  {Schoelkopf}}]{spietz_primary_2003}%
  \BibitemOpen
  \bibfield  {author} {\bibinfo {author} {\bibfnamefont {L.}~\bibnamefont
  {Spietz}}, \bibinfo {author} {\bibfnamefont {K.~W.}\ \bibnamefont {Lehnert}},
  \bibinfo {author} {\bibfnamefont {I.}~\bibnamefont {Siddiqi}}, \ and\
  \bibinfo {author} {\bibfnamefont {R.~J.}\ \bibnamefont {Schoelkopf}},\ }\href
  {\doibase 10.1126/science.1084647} {\bibfield  {journal} {\bibinfo  {journal}
  {Science}\ }\textbf {\bibinfo {volume} {300}},\ \bibinfo {pages} {1929}
  (\bibinfo {year} {2003})}\BibitemShut {NoStop}%
\bibitem [{\citenamefont {Spietz}\ \emph {et~al.}(2006)\citenamefont {Spietz},
  \citenamefont {Schoelkopf},\ and\ \citenamefont {Pari}}]{spietz_shot_2006}%
  \BibitemOpen
  \bibfield  {author} {\bibinfo {author} {\bibfnamefont {L.}~\bibnamefont
  {Spietz}}, \bibinfo {author} {\bibfnamefont {R.~J.}\ \bibnamefont
  {Schoelkopf}}, \ and\ \bibinfo {author} {\bibfnamefont {P.}~\bibnamefont
  {Pari}},\ }\href {\doibase 10.1063/1.2382736} {\bibfield  {journal} {\bibinfo
   {journal} {Appl. Phys. Lett.}\ }\textbf {\bibinfo {volume} {89}},\ \bibinfo
  {pages} {183123} (\bibinfo {year} {2006})}\BibitemShut {NoStop}%
\bibitem [{\citenamefont {Gasparinetti}\ \emph {et~al.}(2011)\citenamefont
  {Gasparinetti}, \citenamefont {Deon}, \citenamefont {Biasiol}, \citenamefont
  {Sorba}, \citenamefont {Beltram},\ and\ \citenamefont
  {Giazotto}}]{gasparinetti2011probing}%
  \BibitemOpen
  \bibfield  {author} {\bibinfo {author} {\bibfnamefont {S.}~\bibnamefont
  {Gasparinetti}}, \bibinfo {author} {\bibfnamefont {F.}~\bibnamefont {Deon}},
  \bibinfo {author} {\bibfnamefont {G.}~\bibnamefont {Biasiol}}, \bibinfo
  {author} {\bibfnamefont {L.}~\bibnamefont {Sorba}}, \bibinfo {author}
  {\bibfnamefont {F.}~\bibnamefont {Beltram}}, \ and\ \bibinfo {author}
  {\bibfnamefont {F.}~\bibnamefont {Giazotto}},\ }\href@noop {} {\bibfield
  {journal} {\bibinfo  {journal} {Physical Review B}\ }\textbf {\bibinfo
  {volume} {83}},\ \bibinfo {pages} {201306} (\bibinfo {year}
  {2011})}\BibitemShut {NoStop}%
\bibitem [{\citenamefont {Mavalankar}\ \emph {et~al.}(2013)\citenamefont
  {Mavalankar}, \citenamefont {Chorley}, \citenamefont {Griffiths},
  \citenamefont {Jones}, \citenamefont {Farrer}, \citenamefont {Ritchie},\ and\
  \citenamefont {Smith}}]{mavalankar_non-invasive_2013}%
  \BibitemOpen
  \bibfield  {author} {\bibinfo {author} {\bibfnamefont {A.}~\bibnamefont
  {Mavalankar}}, \bibinfo {author} {\bibfnamefont {S.~J.}\ \bibnamefont
  {Chorley}}, \bibinfo {author} {\bibfnamefont {J.}~\bibnamefont {Griffiths}},
  \bibinfo {author} {\bibfnamefont {G.~a.~C.}\ \bibnamefont {Jones}}, \bibinfo
  {author} {\bibfnamefont {I.}~\bibnamefont {Farrer}}, \bibinfo {author}
  {\bibfnamefont {D.~A.}\ \bibnamefont {Ritchie}}, \ and\ \bibinfo {author}
  {\bibfnamefont {C.~G.}\ \bibnamefont {Smith}},\ }\href {\doibase
  10.1063/1.4823703} {\bibfield  {journal} {\bibinfo  {journal} {Appl. Phys.
  Lett.}\ }\textbf {\bibinfo {volume} {103}},\ \bibinfo {pages} {133116}
  (\bibinfo {year} {2013})}\BibitemShut {NoStop}%
\bibitem [{\citenamefont {Feshchenko}\ \emph {et~al.}(2013)\citenamefont
  {Feshchenko}, \citenamefont {Meschke}, \citenamefont {Gunnarsson},
  \citenamefont {Prunnila}, \citenamefont {Roschier}, \citenamefont
  {Penttil{\ifmmode\ddot{a}\else\"{a}\fi}},\ and\ \citenamefont
  {Pekola}}]{feshchenko_primary_2013}%
  \BibitemOpen
  \bibfield  {author} {\bibinfo {author} {\bibfnamefont {A.~V.}\ \bibnamefont
  {Feshchenko}}, \bibinfo {author} {\bibfnamefont {M.}~\bibnamefont {Meschke}},
  \bibinfo {author} {\bibfnamefont {D.}~\bibnamefont {Gunnarsson}}, \bibinfo
  {author} {\bibfnamefont {M.}~\bibnamefont {Prunnila}}, \bibinfo {author}
  {\bibfnamefont {L.}~\bibnamefont {Roschier}}, \bibinfo {author}
  {\bibfnamefont {J.~S.}\ \bibnamefont
  {Penttil{\ifmmode\ddot{a}\else\"{a}\fi}}}, \ and\ \bibinfo {author}
  {\bibfnamefont {J.~P.}\ \bibnamefont {Pekola}},\ }\href {\doibase
  10.1007/s10909-013-0874-x} {\bibfield  {journal} {\bibinfo  {journal} {J. Low
  Temp. Phys.}\ }\textbf {\bibinfo {volume} {173}},\ \bibinfo {pages} {36}
  (\bibinfo {year} {2013})}\BibitemShut {NoStop}%
\bibitem [{\citenamefont {Maradan}\ \emph {et~al.}(2014)\citenamefont
  {Maradan}, \citenamefont {Casparis}, \citenamefont {Liu}, \citenamefont
  {Biesinger}, \citenamefont {Scheller}, \citenamefont {Zumbühl},
  \citenamefont {Zimmerman},\ and\ \citenamefont
  {Gossard}}]{maradan_gaas_2014}%
  \BibitemOpen
  \bibfield  {author} {\bibinfo {author} {\bibfnamefont {D.}~\bibnamefont
  {Maradan}}, \bibinfo {author} {\bibfnamefont {L.}~\bibnamefont {Casparis}},
  \bibinfo {author} {\bibfnamefont {T.-M.}\ \bibnamefont {Liu}}, \bibinfo
  {author} {\bibfnamefont {D.~E.~F.}\ \bibnamefont {Biesinger}}, \bibinfo
  {author} {\bibfnamefont {C.~P.}\ \bibnamefont {Scheller}}, \bibinfo {author}
  {\bibfnamefont {D.~M.}\ \bibnamefont {Zumbühl}}, \bibinfo {author}
  {\bibfnamefont {J.~D.}\ \bibnamefont {Zimmerman}}, \ and\ \bibinfo {author}
  {\bibfnamefont {A.~C.}\ \bibnamefont {Gossard}},\ }\href {\doibase
  10.1007/s10909-014-1169-6} {\bibfield  {journal} {\bibinfo  {journal} {J. Low
  Temp. Phys.}\ }\textbf {\bibinfo {volume} {175}},\ \bibinfo {pages} {784}
  (\bibinfo {year} {2014})}\BibitemShut {NoStop}%
\bibitem [{\citenamefont {Feshchenko}\ \emph {et~al.}(2015)\citenamefont
  {Feshchenko}, \citenamefont {Casparis}, \citenamefont {Khaymovich},
  \citenamefont {Maradan}, \citenamefont {Saira}, \citenamefont {Palma},
  \citenamefont {Meschke}, \citenamefont {Pekola},\ and\ \citenamefont
  {Zumb{\ifmmode\ddot{u}\else\"{u}\fi}hl}}]{feshchenko_tunnel-junction_2015}%
  \BibitemOpen
  \bibfield  {author} {\bibinfo {author} {\bibfnamefont {A.~V.}\ \bibnamefont
  {Feshchenko}}, \bibinfo {author} {\bibfnamefont {L.}~\bibnamefont
  {Casparis}}, \bibinfo {author} {\bibfnamefont {I.~M.}\ \bibnamefont
  {Khaymovich}}, \bibinfo {author} {\bibfnamefont {D.}~\bibnamefont {Maradan}},
  \bibinfo {author} {\bibfnamefont {O.-P.}\ \bibnamefont {Saira}}, \bibinfo
  {author} {\bibfnamefont {M.}~\bibnamefont {Palma}}, \bibinfo {author}
  {\bibfnamefont {M.}~\bibnamefont {Meschke}}, \bibinfo {author} {\bibfnamefont
  {J.~P.}\ \bibnamefont {Pekola}}, \ and\ \bibinfo {author} {\bibfnamefont
  {D.~M.}\ \bibnamefont {Zumb{\ifmmode\ddot{u}\else\"{u}\fi}hl}},\ }\href
  {\doibase 10.1103/PhysRevApplied.4.034001} {\bibfield  {journal} {\bibinfo
  {journal} {Phys. Rev. Appl.}\ }\textbf {\bibinfo {volume} {4}},\ \bibinfo
  {pages} {034001} (\bibinfo {year} {2015})}\BibitemShut {NoStop}%
\bibitem [{\citenamefont {Iftikhar}\ \emph {et~al.}(2016)\citenamefont
  {Iftikhar}, \citenamefont {Anthore}, \citenamefont {Jezouin}, \citenamefont
  {Parmentier}, \citenamefont {Jin}, \citenamefont {Cavanna}, \citenamefont
  {Ouerghi}, \citenamefont {Gennser},\ and\ \citenamefont
  {Pierre}}]{iftikhar_primary_2016}%
  \BibitemOpen
  \bibfield  {author} {\bibinfo {author} {\bibfnamefont {Z.}~\bibnamefont
  {Iftikhar}}, \bibinfo {author} {\bibfnamefont {A.}~\bibnamefont {Anthore}},
  \bibinfo {author} {\bibfnamefont {S.}~\bibnamefont {Jezouin}}, \bibinfo
  {author} {\bibfnamefont {F.~D.}\ \bibnamefont {Parmentier}}, \bibinfo
  {author} {\bibfnamefont {Y.}~\bibnamefont {Jin}}, \bibinfo {author}
  {\bibfnamefont {A.}~\bibnamefont {Cavanna}}, \bibinfo {author} {\bibfnamefont
  {A.}~\bibnamefont {Ouerghi}}, \bibinfo {author} {\bibfnamefont
  {U.}~\bibnamefont {Gennser}}, \ and\ \bibinfo {author} {\bibfnamefont
  {F.}~\bibnamefont {Pierre}},\ }\href {\doibase 10.1038/ncomms12908}
  {\bibfield  {journal} {\bibinfo  {journal} {Nat. Commun.}\ }\textbf {\bibinfo
  {volume} {7}},\ \bibinfo {pages} {12908} (\bibinfo {year}
  {2016})}\BibitemShut {NoStop}%
\bibitem [{\citenamefont {Ahmed}\ \emph {et~al.}(2018)\citenamefont {Ahmed},
  \citenamefont {Chatterjee}, \citenamefont {Barraud}, \citenamefont {Morton},
  \citenamefont {Haigh},\ and\ \citenamefont {Gonzalez-Zalba}}]{Ahmed2018Oct}%
  \BibitemOpen
  \bibfield  {author} {\bibinfo {author} {\bibfnamefont {I.}~\bibnamefont
  {Ahmed}}, \bibinfo {author} {\bibfnamefont {A.}~\bibnamefont {Chatterjee}},
  \bibinfo {author} {\bibfnamefont {S.}~\bibnamefont {Barraud}}, \bibinfo
  {author} {\bibfnamefont {J.~J.~L.}\ \bibnamefont {Morton}}, \bibinfo {author}
  {\bibfnamefont {J.~A.}\ \bibnamefont {Haigh}}, \ and\ \bibinfo {author}
  {\bibfnamefont {M.~F.}\ \bibnamefont {Gonzalez-Zalba}},\ }\href {\doibase
  10.1038/s42005-018-0066-8} {\bibfield  {journal} {\bibinfo  {journal}
  {Communications Physics}\ }\textbf {\bibinfo {volume} {1}},\ \bibinfo {pages}
  {66} (\bibinfo {year} {2018})}\BibitemShut {NoStop}%
\bibitem [{\citenamefont {Karimi}\ and\ \citenamefont
  {Pekola}(2018)}]{karimi2018noninvasive}%
  \BibitemOpen
  \bibfield  {author} {\bibinfo {author} {\bibfnamefont {B.}~\bibnamefont
  {Karimi}}\ and\ \bibinfo {author} {\bibfnamefont {J.~P.}\ \bibnamefont
  {Pekola}},\ }\href@noop {} {\bibfield  {journal} {\bibinfo  {journal}
  {Physical Review Applied}\ }\textbf {\bibinfo {volume} {10}},\ \bibinfo
  {pages} {054048} (\bibinfo {year} {2018})}\BibitemShut {NoStop}%
\bibitem [{\citenamefont {Halbertal}\ \emph {et~al.}(2016)\citenamefont
  {Halbertal}, \citenamefont {Cuppens}, \citenamefont {Shalom}, \citenamefont
  {Embon}, \citenamefont {Shadmi}, \citenamefont {Anahory}, \citenamefont
  {Naren}, \citenamefont {Sarkar}, \citenamefont {Uri}, \citenamefont {Ronen}
  \emph {et~al.}}]{halbertal2016nanoscale}%
  \BibitemOpen
  \bibfield  {author} {\bibinfo {author} {\bibfnamefont {D.}~\bibnamefont
  {Halbertal}}, \bibinfo {author} {\bibfnamefont {J.}~\bibnamefont {Cuppens}},
  \bibinfo {author} {\bibfnamefont {M.~B.}\ \bibnamefont {Shalom}}, \bibinfo
  {author} {\bibfnamefont {L.}~\bibnamefont {Embon}}, \bibinfo {author}
  {\bibfnamefont {N.}~\bibnamefont {Shadmi}}, \bibinfo {author} {\bibfnamefont
  {Y.}~\bibnamefont {Anahory}}, \bibinfo {author} {\bibfnamefont
  {H.}~\bibnamefont {Naren}}, \bibinfo {author} {\bibfnamefont
  {J.}~\bibnamefont {Sarkar}}, \bibinfo {author} {\bibfnamefont
  {A.}~\bibnamefont {Uri}}, \bibinfo {author} {\bibfnamefont {Y.}~\bibnamefont
  {Ronen}},  \emph {et~al.},\ }\href@noop {} {\bibfield  {journal} {\bibinfo
  {journal} {Nature}\ }\textbf {\bibinfo {volume} {539}},\ \bibinfo {pages}
  {407} (\bibinfo {year} {2016})}\BibitemShut {NoStop}%
\bibitem [{\citenamefont {Zgirski}\ \emph {et~al.}(2018)\citenamefont
  {Zgirski}, \citenamefont {Foltyn}, \citenamefont {Savin}, \citenamefont
  {Norowski}, \citenamefont {Meschke},\ and\ \citenamefont
  {Pekola}}]{zgirski2018nanosecond}%
  \BibitemOpen
  \bibfield  {author} {\bibinfo {author} {\bibfnamefont {M.}~\bibnamefont
  {Zgirski}}, \bibinfo {author} {\bibfnamefont {M.}~\bibnamefont {Foltyn}},
  \bibinfo {author} {\bibfnamefont {A.}~\bibnamefont {Savin}}, \bibinfo
  {author} {\bibfnamefont {K.}~\bibnamefont {Norowski}}, \bibinfo {author}
  {\bibfnamefont {M.}~\bibnamefont {Meschke}}, \ and\ \bibinfo {author}
  {\bibfnamefont {J.}~\bibnamefont {Pekola}},\ }\href {\doibase
  10.1103/PhysRevApplied.10.044068} {\bibfield  {journal} {\bibinfo  {journal}
  {Phys. Rev. Appl.}\ }\textbf {\bibinfo {volume} {10}},\ \bibinfo {pages}
  {044068} (\bibinfo {year} {2018})}\BibitemShut {NoStop}%
\bibitem [{\citenamefont {Wang}\ \emph {et~al.}(2018)\citenamefont {Wang},
  \citenamefont {Saira},\ and\ \citenamefont {Pekola}}]{wang2018fast}%
  \BibitemOpen
  \bibfield  {author} {\bibinfo {author} {\bibfnamefont {L.}~\bibnamefont
  {Wang}}, \bibinfo {author} {\bibfnamefont {O.-P.}\ \bibnamefont {Saira}}, \
  and\ \bibinfo {author} {\bibfnamefont {J.}~\bibnamefont {Pekola}},\
  }\href@noop {} {\bibfield  {journal} {\bibinfo  {journal} {Applied Physics
  Letters}\ }\textbf {\bibinfo {volume} {112}},\ \bibinfo {pages} {013105}
  (\bibinfo {year} {2018})}\BibitemShut {NoStop}%
\bibitem [{\citenamefont {Brange}\ \emph {et~al.}(2018)\citenamefont {Brange},
  \citenamefont {Samuelsson}, \citenamefont {Karimi},\ and\ \citenamefont
  {Pekola}}]{brange2018nanoscale}%
  \BibitemOpen
  \bibfield  {author} {\bibinfo {author} {\bibfnamefont {F.}~\bibnamefont
  {Brange}}, \bibinfo {author} {\bibfnamefont {P.}~\bibnamefont {Samuelsson}},
  \bibinfo {author} {\bibfnamefont {B.}~\bibnamefont {Karimi}}, \ and\ \bibinfo
  {author} {\bibfnamefont {J.~P.}\ \bibnamefont {Pekola}},\ }\href@noop {}
  {\bibfield  {journal} {\bibinfo  {journal} {Physical Review B}\ }\textbf
  {\bibinfo {volume} {98}},\ \bibinfo {pages} {205414} (\bibinfo {year}
  {2018})}\BibitemShut {NoStop}%
\bibitem [{\citenamefont {Dar\'e}\ and\ \citenamefont
  {Lombardo}(2017)}]{dare_powerful_2017}%
  \BibitemOpen
  \bibfield  {author} {\bibinfo {author} {\bibfnamefont {A.-M.}\ \bibnamefont
  {Dar\'e}}\ and\ \bibinfo {author} {\bibfnamefont {P.}~\bibnamefont
  {Lombardo}},\ }\href {\doibase 10.1103/PhysRevB.96.115414} {\bibfield
  {journal} {\bibinfo  {journal} {Phys. Rev. B}\ }\textbf {\bibinfo {volume}
  {96}},\ \bibinfo {pages} {115414} (\bibinfo {year} {2017})}\BibitemShut
  {NoStop}%
\bibitem [{\citenamefont {Walldorf}\ \emph {et~al.}(2017)\citenamefont
  {Walldorf}, \citenamefont {Jauho},\ and\ \citenamefont
  {Kaasbjerg}}]{walldorf_thermoelectrics_2017}%
  \BibitemOpen
  \bibfield  {author} {\bibinfo {author} {\bibfnamefont {N.}~\bibnamefont
  {Walldorf}}, \bibinfo {author} {\bibfnamefont {A.-P.}\ \bibnamefont {Jauho}},
  \ and\ \bibinfo {author} {\bibfnamefont {K.}~\bibnamefont {Kaasbjerg}},\
  }\href {\doibase 10.1103/PhysRevB.96.115415} {\bibfield  {journal} {\bibinfo
  {journal} {Phys. Rev. B}\ }\textbf {\bibinfo {volume} {96}},\ \bibinfo
  {pages} {115415} (\bibinfo {year} {2017})}\BibitemShut {NoStop}%
\bibitem [{\citenamefont {Strasberg}\ \emph {et~al.}(2018)\citenamefont
  {Strasberg}, \citenamefont {Schaller}, \citenamefont {Schmidt},\ and\
  \citenamefont {Esposito}}]{strasberg_fermionic_2018}%
  \BibitemOpen
  \bibfield  {author} {\bibinfo {author} {\bibfnamefont {P.}~\bibnamefont
  {Strasberg}}, \bibinfo {author} {\bibfnamefont {G.}~\bibnamefont {Schaller}},
  \bibinfo {author} {\bibfnamefont {T.~L.}\ \bibnamefont {Schmidt}}, \ and\
  \bibinfo {author} {\bibfnamefont {M.}~\bibnamefont {Esposito}},\ }\href
  {\doibase 10.1103/PhysRevB.97.205405} {\bibfield  {journal} {\bibinfo
  {journal} {Phys. Rev. B}\ }\textbf {\bibinfo {volume} {97}},\ \bibinfo
  {pages} {205405} (\bibinfo {year} {2018})}\BibitemShut {NoStop}%
\bibitem [{\citenamefont {Erdman}\ \emph {et~al.}(2018)\citenamefont {Erdman},
  \citenamefont {Bhandari}, \citenamefont {Fazio}, \citenamefont {Pekola},\
  and\ \citenamefont {Taddei}}]{erdman_absorption_2018}%
  \BibitemOpen
  \bibfield  {author} {\bibinfo {author} {\bibfnamefont {P.~A.}\ \bibnamefont
  {Erdman}}, \bibinfo {author} {\bibfnamefont {B.}~\bibnamefont {Bhandari}},
  \bibinfo {author} {\bibfnamefont {R.}~\bibnamefont {Fazio}}, \bibinfo
  {author} {\bibfnamefont {J.~P.}\ \bibnamefont {Pekola}}, \ and\ \bibinfo
  {author} {\bibfnamefont {F.}~\bibnamefont {Taddei}},\ }\href {\doibase
  10.1103/PhysRevB.98.045433} {\bibfield  {journal} {\bibinfo  {journal} {Phys.
  Rev. B}\ }\textbf {\bibinfo {volume} {98}},\ \bibinfo {pages} {045433}
  (\bibinfo {year} {2018})}\BibitemShut {NoStop}%
\bibitem [{\citenamefont {Thierschmann}\ \emph
  {et~al.}(2015{\natexlab{b}})\citenamefont {Thierschmann}, \citenamefont
  {Arnold}, \citenamefont {Mittermüller}, \citenamefont {Maier}, \citenamefont
  {Heyn}, \citenamefont {Hansen}, \citenamefont {Buhmann},\ and\ \citenamefont
  {Molenkamp}}]{thierschmann_thermal_2015}%
  \BibitemOpen
  \bibfield  {author} {\bibinfo {author} {\bibfnamefont {H.}~\bibnamefont
  {Thierschmann}}, \bibinfo {author} {\bibfnamefont {F.}~\bibnamefont
  {Arnold}}, \bibinfo {author} {\bibfnamefont {M.}~\bibnamefont
  {Mittermüller}}, \bibinfo {author} {\bibfnamefont {L.}~\bibnamefont
  {Maier}}, \bibinfo {author} {\bibfnamefont {C.}~\bibnamefont {Heyn}},
  \bibinfo {author} {\bibfnamefont {W.}~\bibnamefont {Hansen}}, \bibinfo
  {author} {\bibfnamefont {H.}~\bibnamefont {Buhmann}}, \ and\ \bibinfo
  {author} {\bibfnamefont {L.~W.}\ \bibnamefont {Molenkamp}},\ }\href {\doibase
  10.1088/1367-2630/17/11/113003} {\bibfield  {journal} {\bibinfo  {journal}
  {New J. Phys.}\ }\textbf {\bibinfo {volume} {17}},\ \bibinfo {pages} {113003}
  (\bibinfo {year} {2015}{\natexlab{b}})}\BibitemShut {NoStop}%
\bibitem [{\citenamefont {Zhang}\ and\ \citenamefont
  {Chen}(2018)}]{Zhang2018Nov}%
  \BibitemOpen
  \bibfield  {author} {\bibinfo {author} {\bibfnamefont {Y.}~\bibnamefont
  {Zhang}}\ and\ \bibinfo {author} {\bibfnamefont {J.}~\bibnamefont {Chen}},\
  }\href {https://arxiv.org/abs/1811.11335} {\bibfield  {journal} {\bibinfo
  {journal} {arXiv}\ } (\bibinfo {year} {2018})},\ \Eprint
  {http://arxiv.org/abs/1811.11335} {1811.11335} \BibitemShut {NoStop}%
\bibitem [{\citenamefont {Korotkov}(1999)}]{korotkov_continuous_1999}%
  \BibitemOpen
  \bibfield  {author} {\bibinfo {author} {\bibfnamefont {A.~N.}\ \bibnamefont
  {Korotkov}},\ }\href {\doibase 10.1103/PhysRevB.60.5737} {\bibfield
  {journal} {\bibinfo  {journal} {Phys. Rev. B}\ }\textbf {\bibinfo {volume}
  {60}},\ \bibinfo {pages} {5737} (\bibinfo {year} {1999})}\BibitemShut
  {NoStop}%
\bibitem [{\citenamefont {Korotkov}(2001)}]{korotkov_selective_2001}%
  \BibitemOpen
  \bibfield  {author} {\bibinfo {author} {\bibfnamefont {A.~N.}\ \bibnamefont
  {Korotkov}},\ }\href {\doibase 10.1103/PhysRevB.63.115403} {\bibfield
  {journal} {\bibinfo  {journal} {Phys. Rev. B}\ }\textbf {\bibinfo {volume}
  {63}},\ \bibinfo {pages} {115403} (\bibinfo {year} {2001})}\BibitemShut
  {NoStop}%
\bibitem [{\citenamefont {Korotkov}(2003)}]{korotkov_noisy_2002}%
  \BibitemOpen
  \bibfield  {author} {\bibinfo {author} {\bibfnamefont {A.~N.}\ \bibnamefont
  {Korotkov}},\ }\bibfield  {booktitle} {\emph {\bibinfo {booktitle} {Quantum
  Noise in Mesoscopic Physics}},\ }\href {\doibase
  10.1007/978-94-010-0089-5_10} {\ ,\ \bibinfo {pages} {205} (\bibinfo {year}
  {2003})}\BibitemShut {NoStop}%
\bibitem [{\citenamefont {Pilgram}\ and\ \citenamefont
  {B\"uttiker}(2002)}]{pilgram_efficiency_2002}%
  \BibitemOpen
  \bibfield  {author} {\bibinfo {author} {\bibfnamefont {S.}~\bibnamefont
  {Pilgram}}\ and\ \bibinfo {author} {\bibfnamefont {M.}~\bibnamefont
  {B\"uttiker}},\ }\href {\doibase 10.1103/PhysRevLett.89.200401} {\bibfield
  {journal} {\bibinfo  {journal} {Phys. Rev. Lett.}\ }\textbf {\bibinfo
  {volume} {89}},\ \bibinfo {pages} {200401} (\bibinfo {year}
  {2002})}\BibitemShut {NoStop}%
\bibitem [{\citenamefont {Jordan}\ and\ \citenamefont
  {B\"uttiker}(2005{\natexlab{a}})}]{jordan_continuous_2005}%
  \BibitemOpen
  \bibfield  {author} {\bibinfo {author} {\bibfnamefont {A.~N.}\ \bibnamefont
  {Jordan}}\ and\ \bibinfo {author} {\bibfnamefont {M.}~\bibnamefont
  {B\"uttiker}},\ }\href {\doibase 10.1103/PhysRevLett.95.220401} {\bibfield
  {journal} {\bibinfo  {journal} {Phys. Rev. Lett.}\ }\textbf {\bibinfo
  {volume} {95}},\ \bibinfo {pages} {220401} (\bibinfo {year}
  {2005}{\natexlab{a}})}\BibitemShut {NoStop}%
\bibitem [{\citenamefont {Jordan}\ and\ \citenamefont
  {B\"uttiker}(2005{\natexlab{b}})}]{jordan_quantum_2005}%
  \BibitemOpen
  \bibfield  {author} {\bibinfo {author} {\bibfnamefont {A.~N.}\ \bibnamefont
  {Jordan}}\ and\ \bibinfo {author} {\bibfnamefont {M.}~\bibnamefont
  {B\"uttiker}},\ }\href {\doibase 10.1103/PhysRevB.71.125333} {\bibfield
  {journal} {\bibinfo  {journal} {Phys. Rev. B}\ }\textbf {\bibinfo {volume}
  {71}},\ \bibinfo {pages} {125333} (\bibinfo {year}
  {2005}{\natexlab{b}})}\BibitemShut {NoStop}%
\bibitem [{\citenamefont {Jordan}\ and\ \citenamefont
  {Korotkov}(2006)}]{jordan_qubit_2006}%
  \BibitemOpen
  \bibfield  {author} {\bibinfo {author} {\bibfnamefont {A.~N.}\ \bibnamefont
  {Jordan}}\ and\ \bibinfo {author} {\bibfnamefont {A.~N.}\ \bibnamefont
  {Korotkov}},\ }\href {\doibase 10.1103/PhysRevB.74.085307} {\bibfield
  {journal} {\bibinfo  {journal} {Phys. Rev. B}\ }\textbf {\bibinfo {volume}
  {74}},\ \bibinfo {pages} {085307} (\bibinfo {year} {2006})}\BibitemShut
  {NoStop}%
\bibitem [{\citenamefont {Jordan}\ \emph {et~al.}(2006)\citenamefont {Jordan},
  \citenamefont {Korotkov},\ and\ \citenamefont
  {B\"uttiker}}]{jordan_leggett-garg_2006}%
  \BibitemOpen
  \bibfield  {author} {\bibinfo {author} {\bibfnamefont {A.~N.}\ \bibnamefont
  {Jordan}}, \bibinfo {author} {\bibfnamefont {A.~N.}\ \bibnamefont
  {Korotkov}}, \ and\ \bibinfo {author} {\bibfnamefont {M.}~\bibnamefont
  {B\"uttiker}},\ }\href {\doibase 10.1103/PhysRevLett.97.026805} {\bibfield
  {journal} {\bibinfo  {journal} {Phys. Rev. Lett.}\ }\textbf {\bibinfo
  {volume} {97}},\ \bibinfo {pages} {026805} (\bibinfo {year}
  {2006})}\BibitemShut {NoStop}%
\bibitem [{\citenamefont {Sukhorukov}\ \emph {et~al.}(2007)\citenamefont
  {Sukhorukov}, \citenamefont {Jordan}, \citenamefont {Gustavsson},
  \citenamefont {Leturcq}, \citenamefont {Ihn},\ and\ \citenamefont
  {Ensslin}}]{sukhorukov_conditional_2007}%
  \BibitemOpen
  \bibfield  {author} {\bibinfo {author} {\bibfnamefont {E.~V.}\ \bibnamefont
  {Sukhorukov}}, \bibinfo {author} {\bibfnamefont {A.~N.}\ \bibnamefont
  {Jordan}}, \bibinfo {author} {\bibfnamefont {S.}~\bibnamefont {Gustavsson}},
  \bibinfo {author} {\bibfnamefont {R.}~\bibnamefont {Leturcq}}, \bibinfo
  {author} {\bibfnamefont {T.}~\bibnamefont {Ihn}}, \ and\ \bibinfo {author}
  {\bibfnamefont {K.}~\bibnamefont {Ensslin}},\ }\href {\doibase
  10.1038/nphys564} {\bibfield  {journal} {\bibinfo  {journal} {Nat. Phys.}\
  }\textbf {\bibinfo {volume} {3}},\ \bibinfo {pages} {243} (\bibinfo {year}
  {2007})}\BibitemShut {NoStop}%
\bibitem [{\citenamefont {Flindt}\ \emph {et~al.}(2009)\citenamefont {Flindt},
  \citenamefont {Fricke}, \citenamefont {Hohls}, \citenamefont
  {Novotn{\ifmmode\acute{y}\else\'{y}\fi}}, \citenamefont
  {Neto{\ifmmode\check{c}\else\v{c}\fi}n{\ifmmode\acute{y}\else\'{y}\fi}},
  \citenamefont {Brandes},\ and\ \citenamefont {Haug}}]{flindt_universal_2009}%
  \BibitemOpen
  \bibfield  {author} {\bibinfo {author} {\bibfnamefont {C.}~\bibnamefont
  {Flindt}}, \bibinfo {author} {\bibfnamefont {C.}~\bibnamefont {Fricke}},
  \bibinfo {author} {\bibfnamefont {F.}~\bibnamefont {Hohls}}, \bibinfo
  {author} {\bibfnamefont {T.}~\bibnamefont
  {Novotn{\ifmmode\acute{y}\else\'{y}\fi}}}, \bibinfo {author} {\bibfnamefont
  {K.}~\bibnamefont
  {Neto{\ifmmode\check{c}\else\v{c}\fi}n{\ifmmode\acute{y}\else\'{y}\fi}}},
  \bibinfo {author} {\bibfnamefont {T.}~\bibnamefont {Brandes}}, \ and\
  \bibinfo {author} {\bibfnamefont {R.~J.}\ \bibnamefont {Haug}},\ }\href
  {\doibase 10.1073/pnas.0901002106} {\bibfield  {journal} {\bibinfo  {journal}
  {Proc. Natl. Acad. Sci. U.S.A.}\ }\textbf {\bibinfo {volume} {106}},\
  \bibinfo {pages} {10116} (\bibinfo {year} {2009})}\BibitemShut {NoStop}%
\bibitem [{\citenamefont {Gustavsson}\ \emph {et~al.}(2006)\citenamefont
  {Gustavsson}, \citenamefont {Leturcq}, \citenamefont {Simovič},
  \citenamefont {Schleser}, \citenamefont {Ihn}, \citenamefont {Studerus},
  \citenamefont {Ensslin}, \citenamefont {Driscoll},\ and\ \citenamefont
  {Gossard}}]{gustavsson_counting_2006}%
  \BibitemOpen
  \bibfield  {author} {\bibinfo {author} {\bibfnamefont {S.}~\bibnamefont
  {Gustavsson}}, \bibinfo {author} {\bibfnamefont {R.}~\bibnamefont {Leturcq}},
  \bibinfo {author} {\bibfnamefont {B.}~\bibnamefont {Simovič}}, \bibinfo
  {author} {\bibfnamefont {R.}~\bibnamefont {Schleser}}, \bibinfo {author}
  {\bibfnamefont {T.}~\bibnamefont {Ihn}}, \bibinfo {author} {\bibfnamefont
  {P.}~\bibnamefont {Studerus}}, \bibinfo {author} {\bibfnamefont
  {K.}~\bibnamefont {Ensslin}}, \bibinfo {author} {\bibfnamefont {D.~C.}\
  \bibnamefont {Driscoll}}, \ and\ \bibinfo {author} {\bibfnamefont {A.~C.}\
  \bibnamefont {Gossard}},\ }\href {\doibase 10.1103/PhysRevLett.96.076605}
  {\bibfield  {journal} {\bibinfo  {journal} {Phys. Rev. Lett.}\ }\textbf
  {\bibinfo {volume} {96}},\ \bibinfo {pages} {076605} (\bibinfo {year}
  {2006})}\BibitemShut {NoStop}%
\bibitem [{\citenamefont {Fujisawa}\ \emph {et~al.}(2006)\citenamefont
  {Fujisawa}, \citenamefont {Hayashi}, \citenamefont {Tomita},\ and\
  \citenamefont {Hirayama}}]{fujisawa_bidirectional_2006}%
  \BibitemOpen
  \bibfield  {author} {\bibinfo {author} {\bibfnamefont {T.}~\bibnamefont
  {Fujisawa}}, \bibinfo {author} {\bibfnamefont {T.}~\bibnamefont {Hayashi}},
  \bibinfo {author} {\bibfnamefont {R.}~\bibnamefont {Tomita}}, \ and\ \bibinfo
  {author} {\bibfnamefont {Y.}~\bibnamefont {Hirayama}},\ }\href {\doibase
  10.1126/science.1126788} {\bibfield  {journal} {\bibinfo  {journal}
  {Science}\ }\textbf {\bibinfo {volume} {312}},\ \bibinfo {pages} {1634}
  (\bibinfo {year} {2006})}\BibitemShut {NoStop}%
\bibitem [{\citenamefont {Ubbelohde}\ \emph {et~al.}(2012)\citenamefont
  {Ubbelohde}, \citenamefont {Fricke}, \citenamefont {Flindt}, \citenamefont
  {Hohls},\ and\ \citenamefont {Haug}}]{ubbelohde_measurement_2012}%
  \BibitemOpen
  \bibfield  {author} {\bibinfo {author} {\bibfnamefont {N.}~\bibnamefont
  {Ubbelohde}}, \bibinfo {author} {\bibfnamefont {C.}~\bibnamefont {Fricke}},
  \bibinfo {author} {\bibfnamefont {C.}~\bibnamefont {Flindt}}, \bibinfo
  {author} {\bibfnamefont {F.}~\bibnamefont {Hohls}}, \ and\ \bibinfo {author}
  {\bibfnamefont {R.~J.}\ \bibnamefont {Haug}},\ }\href {\doibase
  10.1038/ncomms1620} {\bibfield  {journal} {\bibinfo  {journal} {Nat.
  Commun.}\ }\textbf {\bibinfo {volume} {3}},\ \bibinfo {pages} {612} (\bibinfo
  {year} {2012})}\BibitemShut {NoStop}%
\bibitem [{\citenamefont {K\"ung}\ \emph {et~al.}(2012)\citenamefont {K\"ung},
  \citenamefont {R\"ossler}, \citenamefont {Beck}, \citenamefont {Marthaler},
  \citenamefont {Golubev}, \citenamefont {Utsumi}, \citenamefont {Ihn},\ and\
  \citenamefont {Ensslin}}]{kung_irreversibility_2012}%
  \BibitemOpen
  \bibfield  {author} {\bibinfo {author} {\bibfnamefont {B.}~\bibnamefont
  {K\"ung}}, \bibinfo {author} {\bibfnamefont {C.}~\bibnamefont {R\"ossler}},
  \bibinfo {author} {\bibfnamefont {M.}~\bibnamefont {Beck}}, \bibinfo {author}
  {\bibfnamefont {M.}~\bibnamefont {Marthaler}}, \bibinfo {author}
  {\bibfnamefont {D.~S.}\ \bibnamefont {Golubev}}, \bibinfo {author}
  {\bibfnamefont {Y.}~\bibnamefont {Utsumi}}, \bibinfo {author} {\bibfnamefont
  {T.}~\bibnamefont {Ihn}}, \ and\ \bibinfo {author} {\bibfnamefont
  {K.}~\bibnamefont {Ensslin}},\ }\href {\doibase 10.1103/PhysRevX.2.011001}
  {\bibfield  {journal} {\bibinfo  {journal} {Phys. Rev. X}\ }\textbf {\bibinfo
  {volume} {2}},\ \bibinfo {pages} {011001} (\bibinfo {year}
  {2012})}\BibitemShut {NoStop}%
\bibitem [{\citenamefont {Hofmann}\ \emph
  {et~al.}(2016{\natexlab{a}})\citenamefont {Hofmann}, \citenamefont {Maisi},
  \citenamefont {R{\ifmmode\ddot{o}\else\"{o}\fi}ssler}, \citenamefont
  {Basset}, \citenamefont {Kr{\ifmmode\ddot{a}\else\"{a}\fi}henmann},
  \citenamefont {M{\ifmmode\ddot{a}\else\"{a}\fi}rki}, \citenamefont {Ihn},
  \citenamefont {Ensslin}, \citenamefont {Reichl},\ and\ \citenamefont
  {Wegscheider}}]{hofmann_equilibrium_2016}%
  \BibitemOpen
  \bibfield  {author} {\bibinfo {author} {\bibfnamefont {A.}~\bibnamefont
  {Hofmann}}, \bibinfo {author} {\bibfnamefont {V.~F.}\ \bibnamefont {Maisi}},
  \bibinfo {author} {\bibfnamefont {C.}~\bibnamefont
  {R{\ifmmode\ddot{o}\else\"{o}\fi}ssler}}, \bibinfo {author} {\bibfnamefont
  {J.}~\bibnamefont {Basset}}, \bibinfo {author} {\bibfnamefont
  {T.}~\bibnamefont {Kr{\ifmmode\ddot{a}\else\"{a}\fi}henmann}}, \bibinfo
  {author} {\bibfnamefont {P.}~\bibnamefont
  {M{\ifmmode\ddot{a}\else\"{a}\fi}rki}}, \bibinfo {author} {\bibfnamefont
  {T.}~\bibnamefont {Ihn}}, \bibinfo {author} {\bibfnamefont {K.}~\bibnamefont
  {Ensslin}}, \bibinfo {author} {\bibfnamefont {C.}~\bibnamefont {Reichl}}, \
  and\ \bibinfo {author} {\bibfnamefont {W.}~\bibnamefont {Wegscheider}},\
  }\href {\doibase 10.1103/PhysRevB.93.035425} {\bibfield  {journal} {\bibinfo
  {journal} {Phys. Rev. B}\ }\textbf {\bibinfo {volume} {93}},\ \bibinfo
  {pages} {035425} (\bibinfo {year} {2016}{\natexlab{a}})}\BibitemShut
  {NoStop}%
\bibitem [{\citenamefont {Hofmann}\ \emph
  {et~al.}(2016{\natexlab{b}})\citenamefont {Hofmann}, \citenamefont {Maisi},
  \citenamefont {Gold}, \citenamefont
  {Kr{\ifmmode\ddot{a}\else\"{a}\fi}henmann}, \citenamefont
  {R{\ifmmode\ddot{o}\else\"{o}\fi}ssler}, \citenamefont {Basset},
  \citenamefont {M{\ifmmode\ddot{a}\else\"{a}\fi}rki}, \citenamefont {Reichl},
  \citenamefont {Wegscheider}, \citenamefont {Ensslin},\ and\ \citenamefont
  {Ihn}}]{hofmann_measuring_2016}%
  \BibitemOpen
  \bibfield  {author} {\bibinfo {author} {\bibfnamefont {A.}~\bibnamefont
  {Hofmann}}, \bibinfo {author} {\bibfnamefont {V.~F.}\ \bibnamefont {Maisi}},
  \bibinfo {author} {\bibfnamefont {C.}~\bibnamefont {Gold}}, \bibinfo {author}
  {\bibfnamefont {T.}~\bibnamefont {Kr{\ifmmode\ddot{a}\else\"{a}\fi}henmann}},
  \bibinfo {author} {\bibfnamefont {C.}~\bibnamefont
  {R{\ifmmode\ddot{o}\else\"{o}\fi}ssler}}, \bibinfo {author} {\bibfnamefont
  {J.}~\bibnamefont {Basset}}, \bibinfo {author} {\bibfnamefont
  {P.}~\bibnamefont {M{\ifmmode\ddot{a}\else\"{a}\fi}rki}}, \bibinfo {author}
  {\bibfnamefont {C.}~\bibnamefont {Reichl}}, \bibinfo {author} {\bibfnamefont
  {W.}~\bibnamefont {Wegscheider}}, \bibinfo {author} {\bibfnamefont
  {K.}~\bibnamefont {Ensslin}}, \ and\ \bibinfo {author} {\bibfnamefont
  {T.}~\bibnamefont {Ihn}},\ }\href {\doibase 10.1103/PhysRevLett.117.206803}
  {\bibfield  {journal} {\bibinfo  {journal} {Phys. Rev. Lett.}\ }\textbf
  {\bibinfo {volume} {117}},\ \bibinfo {pages} {206803} (\bibinfo {year}
  {2016}{\natexlab{b}})}\BibitemShut {NoStop}%
\bibitem [{\citenamefont {Entin-Wohlman}\ \emph {et~al.}(2017)\citenamefont
  {Entin-Wohlman}, \citenamefont {Chowdhury}, \citenamefont {Aharony},\ and\
  \citenamefont {Dattagupta}}]{entin-wohlman_heat_2017}%
  \BibitemOpen
  \bibfield  {author} {\bibinfo {author} {\bibfnamefont {O.}~\bibnamefont
  {Entin-Wohlman}}, \bibinfo {author} {\bibfnamefont {D.}~\bibnamefont
  {Chowdhury}}, \bibinfo {author} {\bibfnamefont {A.}~\bibnamefont {Aharony}},
  \ and\ \bibinfo {author} {\bibfnamefont {S.}~\bibnamefont {Dattagupta}},\
  }\href {\doibase 10.1103/PhysRevB.96.195435} {\bibfield  {journal} {\bibinfo
  {journal} {Phys. Rev. B}\ }\textbf {\bibinfo {volume} {96}},\ \bibinfo
  {pages} {195435} (\bibinfo {year} {2017})}\BibitemShut {NoStop}%
\bibitem [{\citenamefont {Gasparinetti}\ \emph {et~al.}(2012)\citenamefont
  {Gasparinetti}, \citenamefont {Mart{\'\i}nez-P{\'e}rez}, \citenamefont
  {De~Franceschi}, \citenamefont {Pekola},\ and\ \citenamefont
  {Giazotto}}]{gasparinetti2012nongalvanic}%
  \BibitemOpen
  \bibfield  {author} {\bibinfo {author} {\bibfnamefont {S.}~\bibnamefont
  {Gasparinetti}}, \bibinfo {author} {\bibfnamefont {M.}~\bibnamefont
  {Mart{\'\i}nez-P{\'e}rez}}, \bibinfo {author} {\bibfnamefont
  {S.}~\bibnamefont {De~Franceschi}}, \bibinfo {author} {\bibfnamefont {J.~P.}\
  \bibnamefont {Pekola}}, \ and\ \bibinfo {author} {\bibfnamefont
  {F.}~\bibnamefont {Giazotto}},\ }\href@noop {} {\bibfield  {journal}
  {\bibinfo  {journal} {Applied Physics Letters}\ }\textbf {\bibinfo {volume}
  {100}},\ \bibinfo {pages} {253502} (\bibinfo {year} {2012})}\BibitemShut
  {NoStop}%
\bibitem [{\citenamefont {Torresani}\ \emph {et~al.}(2013)\citenamefont
  {Torresani}, \citenamefont {Martínez-Pérez}, \citenamefont {Gasparinetti},
  \citenamefont {Renard}, \citenamefont {Biasiol}, \citenamefont {Sorba},
  \citenamefont {Giazotto},\ and\ \citenamefont
  {De~Franceschi}}]{torresani_nongalvanic_2013}%
  \BibitemOpen
  \bibfield  {author} {\bibinfo {author} {\bibfnamefont {P.}~\bibnamefont
  {Torresani}}, \bibinfo {author} {\bibfnamefont {M.~J.}\ \bibnamefont
  {Martínez-Pérez}}, \bibinfo {author} {\bibfnamefont {S.}~\bibnamefont
  {Gasparinetti}}, \bibinfo {author} {\bibfnamefont {J.}~\bibnamefont
  {Renard}}, \bibinfo {author} {\bibfnamefont {G.}~\bibnamefont {Biasiol}},
  \bibinfo {author} {\bibfnamefont {L.}~\bibnamefont {Sorba}}, \bibinfo
  {author} {\bibfnamefont {F.}~\bibnamefont {Giazotto}}, \ and\ \bibinfo
  {author} {\bibfnamefont {S.}~\bibnamefont {De~Franceschi}},\ }\href {\doibase
  10.1103/PhysRevB.88.245304} {\bibfield  {journal} {\bibinfo  {journal} {Phys.
  Rev. B}\ }\textbf {\bibinfo {volume} {88}},\ \bibinfo {pages} {245304}
  (\bibinfo {year} {2013})}\BibitemShut {NoStop}%
\bibitem [{\citenamefont {Giovannetti}\ \emph {et~al.}(2011)\citenamefont
  {Giovannetti}, \citenamefont {Lloyd},\ and\ \citenamefont
  {Maccone}}]{giovannetti_advances_2011}%
  \BibitemOpen
  \bibfield  {author} {\bibinfo {author} {\bibfnamefont {V.}~\bibnamefont
  {Giovannetti}}, \bibinfo {author} {\bibfnamefont {S.}~\bibnamefont {Lloyd}},
  \ and\ \bibinfo {author} {\bibfnamefont {L.}~\bibnamefont {Maccone}},\ }\href
  {\doibase 10.1038/nphoton.2011.35} {\bibfield  {journal} {\bibinfo  {journal}
  {Nat. Photonics}\ }\textbf {\bibinfo {volume} {5}},\ \bibinfo {pages} {222}
  (\bibinfo {year} {2011})}\BibitemShut {NoStop}%
\bibitem [{\citenamefont {Nazarov}\ and\ \citenamefont
  {Blanter}(2009)}]{nazarov_quantum_2009}%
  \BibitemOpen
  \bibfield  {author} {\bibinfo {author} {\bibfnamefont {Y.~V.}\ \bibnamefont
  {Nazarov}}\ and\ \bibinfo {author} {\bibfnamefont {Y.~M.}\ \bibnamefont
  {Blanter}},\ }\href@noop {} {\emph {\bibinfo {title} {Quantum {Transport}:
  {Introduction} to {Nanoscience}}}}\ (\bibinfo  {publisher} {Cambridge
  University Press},\ \bibinfo {address} {Cambridge, UK ; New York},\ \bibinfo
  {year} {2009})\BibitemShut {NoStop}%
\bibitem [{\citenamefont {Ruokola}\ and\ \citenamefont
  {Ojanen}(2012)}]{ruokola_theory_2012}%
  \BibitemOpen
  \bibfield  {author} {\bibinfo {author} {\bibfnamefont {T.}~\bibnamefont
  {Ruokola}}\ and\ \bibinfo {author} {\bibfnamefont {T.}~\bibnamefont
  {Ojanen}},\ }\href {\doibase 10.1103/PhysRevB.86.035454} {\bibfield
  {journal} {\bibinfo  {journal} {Phys. Rev. B}\ }\textbf {\bibinfo {volume}
  {86}},\ \bibinfo {pages} {035454} (\bibinfo {year} {2012})}\BibitemShut
  {NoStop}%
\bibitem [{\citenamefont {Rossell\'o}\ \emph {et~al.}(2017)\citenamefont
  {Rossell\'o}, \citenamefont {L\'opez},\ and\ \citenamefont
  {S\'anchez}}]{rossello_dynamical_2017}%
  \BibitemOpen
  \bibfield  {author} {\bibinfo {author} {\bibfnamefont {G.}~\bibnamefont
  {Rossell\'o}}, \bibinfo {author} {\bibfnamefont {R.}~\bibnamefont {L\'opez}},
  \ and\ \bibinfo {author} {\bibfnamefont {R.}~\bibnamefont {S\'anchez}},\
  }\href {\doibase 10.1103/PhysRevB.95.235404} {\bibfield  {journal} {\bibinfo
  {journal} {Phys. Rev. B}\ }\textbf {\bibinfo {volume} {95}},\ \bibinfo
  {pages} {235404} (\bibinfo {year} {2017})}\BibitemShut {NoStop}%
\bibitem [{\citenamefont {B\"uttiker}(1990)}]{buttiker_quantized_1990}%
  \BibitemOpen
  \bibfield  {author} {\bibinfo {author} {\bibfnamefont {M.}~\bibnamefont
  {B\"uttiker}},\ }\href {\doibase 10.1103/PhysRevB.41.7906} {\bibfield
  {journal} {\bibinfo  {journal} {Phys. Rev. B}\ }\textbf {\bibinfo {volume}
  {41}},\ \bibinfo {pages} {7906} (\bibinfo {year} {1990})}\BibitemShut
  {NoStop}%
\bibitem [{\citenamefont {Christen}\ and\ \citenamefont
  {B\"uttiker}(1996)}]{christen_gauge-invariant_1996}%
  \BibitemOpen
  \bibfield  {author} {\bibinfo {author} {\bibfnamefont {T.}~\bibnamefont
  {Christen}}\ and\ \bibinfo {author} {\bibfnamefont {M.}~\bibnamefont
  {B\"uttiker}},\ }\href {\doibase 10.1209/epl/i1996-00145-8} {\bibfield
  {journal} {\bibinfo  {journal} {Europhysics Letters (EPL)}\ }\textbf
  {\bibinfo {volume} {35}},\ \bibinfo {pages} {523} (\bibinfo {year}
  {1996})}\BibitemShut {NoStop}%
\bibitem [{\citenamefont {Datta}(1997)}]{datta_electronic_1997}%
  \BibitemOpen
  \bibfield  {author} {\bibinfo {author} {\bibfnamefont {S.}~\bibnamefont
  {Datta}},\ }\href@noop {} {\emph {\bibinfo {title} {Electronic {Transport} in
  {Mesoscopic} {Systems}}}}\ (\bibinfo  {publisher} {Cambridge University
  Press},\ \bibinfo {year} {1997})\BibitemShut {NoStop}%
\bibitem [{\citenamefont {Blanter}\ and\ \citenamefont
  {B\"uttiker}(2000)}]{blanter_shot_2000}%
  \BibitemOpen
  \bibfield  {author} {\bibinfo {author} {\bibfnamefont {Y.~M.}\ \bibnamefont
  {Blanter}}\ and\ \bibinfo {author} {\bibfnamefont {M.}~\bibnamefont
  {B\"uttiker}},\ }\href {\doibase 10.1016/S0370-1573(99)00123-4} {\bibfield
  {journal} {\bibinfo  {journal} {Phys. Rep.}\ }\textbf {\bibinfo {volume}
  {336}},\ \bibinfo {pages} {1} (\bibinfo {year} {2000})}\BibitemShut {NoStop}%
\bibitem [{\citenamefont {Kay}(1993)}]{kay_fundamentals_1993}%
  \BibitemOpen
  \bibfield  {author} {\bibinfo {author} {\bibfnamefont {S.~M.}\ \bibnamefont
  {Kay}},\ }\href@noop {} {\emph {\bibinfo {title} {Fundamentals of
  {Statistical} {Processing}, {Volume} {I}: {Estimation} {Theory}}}}\ (\bibinfo
   {publisher} {Prentice Hall},\ \bibinfo {address} {Englewood Cliffs, N.J.},\
  \bibinfo {year} {1993})\BibitemShut {NoStop}%
\bibitem [{\citenamefont {Levitov}\ \emph {et~al.}(1996)\citenamefont
  {Levitov}, \citenamefont {Lee},\ and\ \citenamefont
  {Lesovik}}]{levitov_electron_1996}%
  \BibitemOpen
  \bibfield  {author} {\bibinfo {author} {\bibfnamefont {L.~S.}\ \bibnamefont
  {Levitov}}, \bibinfo {author} {\bibfnamefont {H.}~\bibnamefont {Lee}}, \ and\
  \bibinfo {author} {\bibfnamefont {G.~B.}\ \bibnamefont {Lesovik}},\ }\href
  {\doibase 10.1063/1.531672} {\bibfield  {journal} {\bibinfo  {journal} {J.
  Math. Phys.}\ }\textbf {\bibinfo {volume} {37}},\ \bibinfo {pages} {4845}
  (\bibinfo {year} {1996})}\BibitemShut {NoStop}%
\bibitem [{\citenamefont {Jordan}\ and\ \citenamefont
  {Sukhorukov}(2004)}]{jordan_transport_2004}%
  \BibitemOpen
  \bibfield  {author} {\bibinfo {author} {\bibfnamefont {A.~N.}\ \bibnamefont
  {Jordan}}\ and\ \bibinfo {author} {\bibfnamefont {E.~V.}\ \bibnamefont
  {Sukhorukov}},\ }\href {\doibase 10.1103/PhysRevLett.93.260604} {\bibfield
  {journal} {\bibinfo  {journal} {Phys. Rev. Lett.}\ }\textbf {\bibinfo
  {volume} {93}},\ \bibinfo {pages} {260604} (\bibinfo {year}
  {2004})}\BibitemShut {NoStop}%
\bibitem [{\citenamefont {Singh}\ \emph {et~al.}(2016)\citenamefont {Singh},
  \citenamefont {Peltonen}, \citenamefont {Khaymovich}, \citenamefont {Koski},
  \citenamefont {Flindt},\ and\ \citenamefont
  {Pekola}}]{singh_distribution_2016}%
  \BibitemOpen
  \bibfield  {author} {\bibinfo {author} {\bibfnamefont {S.}~\bibnamefont
  {Singh}}, \bibinfo {author} {\bibfnamefont {J.~T.}\ \bibnamefont {Peltonen}},
  \bibinfo {author} {\bibfnamefont {I.~M.}\ \bibnamefont {Khaymovich}},
  \bibinfo {author} {\bibfnamefont {J.~V.}\ \bibnamefont {Koski}}, \bibinfo
  {author} {\bibfnamefont {C.}~\bibnamefont {Flindt}}, \ and\ \bibinfo {author}
  {\bibfnamefont {J.~P.}\ \bibnamefont {Pekola}},\ }\href {\doibase
  10.1103/PhysRevB.94.241407} {\bibfield  {journal} {\bibinfo  {journal} {Phys.
  Rev. B}\ }\textbf {\bibinfo {volume} {94}},\ \bibinfo {pages} {241407}
  (\bibinfo {year} {2016})}\BibitemShut {NoStop}%
\end{thebibliography}%

\end{document}